\documentclass[journal]{IEEEtran}
\ifCLASSINFOpdf
\else
   \usepackage[dvips]{graphicx}
\fi
\usepackage{url}

\usepackage{booktabs}
\usepackage{mathrsfs}
\usepackage{array}
\usepackage{graphicx}
\usepackage{latexsym}
\usepackage{amsmath}
\usepackage{amsthm}
\usepackage{color}
\usepackage{xcolor}
\usepackage{amsfonts}
\usepackage{subfigure}
\usepackage{dsfont}
\usepackage{epstopdf}
\usepackage{threeparttable}
\usepackage{multirow}
\usepackage{enumerate}
\usepackage{epsfig}
\usepackage{bm}
\usepackage{cite}
\usepackage{algorithmic}
\usepackage{textcomp}
\usepackage{graphicx}
\usepackage{latexsym}
\usepackage{amsmath}
\usepackage{amsthm}
\usepackage{makecell,rotating,multirow,diagbox}
\usepackage{amsfonts}
\usepackage{subfigure}
\usepackage{dsfont}
\usepackage{epstopdf}
\usepackage{threeparttable}
\usepackage{multirow}
\usepackage{enumerate}
\usepackage{epsfig}
\usepackage{bm}
\usepackage{cite}
\usepackage{algorithmic}
\usepackage{textcomp}
\usepackage{xcolor}
\usepackage{amsmath}
\usepackage{cite}

\usepackage{tabu}
\usepackage{makecell}

\makeatletter
\renewcommand{\maketag@@@}[1]{\hbox{\m@th\normalsize\normalfont#1}}%
\makeatother

\ifCLASSINFOpdf
\else
\fi
\begin{document}
%
\title{Deep Learning for 1-Bit Compressed Sensing-based Superimposed CSI Feedback}
%
%
%
\author{{Chaojin Qing\textsuperscript{1*},
Qing Ye\textsuperscript{1},
Bin Cai\textsuperscript{1},
Wenhui Liu\textsuperscript{1},
Jiafan Wang\textsuperscript{2*},}
\\
\bigskip
\textbf{1} School of Electrical Engineering and Electronic Information, Xihua University, Chengdu, China
\\
\textbf{2} Synopsys Inc., 2025 NE Cornelius Pass Rd, Hillsboro, OR 97124, USA
\\
* qingchj@mail.xhu.edu.cn (C. Qing); jifanw@gmail.com(J. Wang)


\thanks{This work is supported in part by the Sichuan Science and Technology Program (Grant No. 2021JDRC0003), the Major Special Funds of Science and Technology of Sichuan Science and Technology Plan Project (Grant No. 19ZDZX0016 /2019YFG0395), the Demonstration Project of Chengdu Major Science and Technology Application (Grant No. 2020-YF09- 00048-SN), the Key Scientific Research Fund of Xihua University (Grant No. Z1120941), and the Special Funds of Industry Development of Sichuan Province (Grant No. zyf-2018-056).}}
\maketitle

\begin{abstract}
In frequency-division duplexing (FDD) massive multiple-input multiple-output (MIMO) systems, 1-bit compressed sensing (CS)-based superimposed channel state information (CSI) feedback has shown many advantages, while still faces many challenges, such as low accuracy of the downlink CSI recovery and large processing delays.
To overcome these drawbacks, this paper proposes a deep learning (DL) scheme to improve the 1-bit compressed sensing-based superimposed CSI feedback.
On the user side, the downlink CSI is compressed with the 1-bit CS technique, superimposed on the uplink user data sequences (UL-US), and then sent back to the base station (BS).
At the BS, based on the model-driven approach and assisted by the superimposition-interference cancellation technology, a multi-task detection network is first constructed for detecting both the UL-US and downlink CSI. In particular, this detection network is jointly trained to detect the UL-US and downlink CSI simultaneously, capturing a globally optimized network parameter.
Then, with the recovered bits for the downlink CSI, a lightweight reconstruction scheme, which consists of an initial feature extraction of the downlink CSI with the simplified traditional method and a single hidden layer network, is utilized to reconstruct the downlink CSI with low processing delay.
Compared with the 1-bit CS-based superimposed CSI feedback scheme, the proposed scheme improves the recovery accuracy of the UL-US and downlink CSI with lower processing delay and possesses robustness against parameter variations.

\end{abstract}


\IEEEpeerreviewmaketitle

\section{Introduction}

\IEEEPARstart{M}{assive} multiple-input multiple-output (MIMO) has become the key technology of the fifth generation (5G) wireless communication system, due to its advantages in system capacity and link robustness~\cite{a1}~\cite{w1}, etc.
As premises of these advantages, the base station (BS) needs to obtain accurate downlink channel state information (CSI), and rely on downlink CSI for precoding~\cite{z1}, antenna selection~\cite{z2}, radio resource allocation~\cite{z3}, and communication interference management~\cite{z4}, etc.
In time division duplex (TDD) mode, the downlink CSI can be obtained from uplink CSI by exploiting channel reciprocity~\cite{a2}~\cite{a3}.
For frequency-division duplex (FDD) mode, it is difficult to develop the channel reciprocity due to the different frequency bands used by uplink and downlink~\cite{a4}\cite{y1}.
Thus, the downlink CSI is usually estimated by users and fed back to the BS in FDD massive MIMO system~\cite{a4}. However, due to a large number of antennas in massive MIMO systems, CSI feedback incurs significant feedback overhead, resulting in serious uplink bandwidth occupation.

To reduce feedback overhead, lots of compressive sensing (CS)-based CSI feedback methods have emerged~\cite{a5,a6,a7,a8}.
In recent years, deep learning (DL)-based CSI feedback methods~\cite{a9,a10,a11} are proposed to further reduce feedback overhead.
Although the feedback overhead is reduced to some extent, both CS-based CSI feedback and DL-based CSI feedback still occupy significant uplink bandwidth resources.
To avoid the occupation of uplink bandwidth resources, the superimposed CSI feedback was proposed in~\cite{a12}, yet causes mutual interference due to superimposition operation. In~\cite{y1},~\cite{b1}, and~\cite{b2}, the 1-bit CS-based, DL-based, and extreme learning machine (ELM)-based superimposed CSI feedbacks are respectively proposed to reduce this mutual interference. Inspired by the advantages of superimposed CSI feedback based on 1-bit CS and DL, we propose a DL-based 1-bit superimposed CSI feedback scheme in this paper.

\subsection{Related Works}

In FDD massive MIMO system, the DL-based CSI feedback methods have been investigated according to the superimposed CSI feedback, e.g., ~\cite{y1},~\cite{b1}, and~\cite{b2}, and feedback reduction, e.g., ~\cite{a13,a14,a15,a16,a17,a18}, etc.

For reducing feedback overhead, the DL-based data-driven CSI feedback can be divided into two categories.
The first category is mainly based on the combination of CS technique and DL technique, while the other category employs the DL technique for the quantized data.
In the first category,~\cite{a13} is the first application of DL for CSI feedback.
In~\cite{a13}, the CSI feedback was mainly based on a convolutional neural network called CsiNet, which achieved superior performance over various CS-based CSI feedbacks.
Yet, the time correlation, frequency correlation, spatial correlation, feedback delay and feedback errors, etc., were not considered in CsiNet, and led to limited applications.
To remedy these defects, some improvements have been proposed in~\cite{a14,a15,a16}.
In~\cite{a14}, a CsiNet long short-term memory (CsiNet-LSTM) was proposed by exploiting the time correlation, which is suitable for practical application in time-varying channels.
The recurrent neural network-based CsiNet in~\cite{a15} was developed to capture the temporal and frequency correlations of wireless channels. Considering the spatial correlation among antennas, the bidirectional LSTM (Bi-LSTM) and bidirectional convolutional LSTM (Bi-ConvLSTM) were proposed in~\cite{a16}. Another category of feedback reduction proposed for DL-based CSI feedback is mainly based on the quantization operation, e.g.,~\cite{a17} and~\cite{a18}.
In~\cite{a17}, a bit-level CsiNet+ was proposed, which made the current CSI feedback network applicable in real communication systems and minimized the introduced quantitative distortion to improve the reconstruction quality.
By employing the quantization and entropy coding blocks into a full convolution network, the work of~\cite{a18} obtained drastic improvement in CSI reconstruction quality at even extremely low feedback rates.
Although the DL-based CSI feedback in~\cite{a13,a14,a15,a16,a17,a18} has achieved significant improvements in feedback reduction compared with the CS-based approaches, the uplink bandwidth resources were still seriously occupied due to the massive MIMO scenarios.

To avoid the occupation of uplink bandwidth resources, superimposed CSI feedback schemes were proposed in~\cite{a12,b1,b2}.
In~\cite{a12}, the downlink CSI was spread and then superimposed on the uplink user data sequences (UL-US) as feedback to the BS, while the recoveries of the UL-US and downlink CSI were deteriorated by superimposition interference.
To remedy this defect, a DL-based superimposed CSI feedback was proposed in~\cite{b1}, and an ELM-based superimposed CSI feedback with lower computational complexity was proposed in~\cite{b2}.
Considering the simplicity and cost-effectiveness, a low-consumed CSI feedback using 1-bit CS has been studied in~\cite{a19}, in which 1-bit operation means to discard the signal amplitude and only retain its sign information.
In this work, the downlink CSI was quantified by 1-bit CS to achieve low-consumed feedback, while this work still occupied uplink bandwidth resources.
To remedy this defect, the superimposed CSI feedback and 1-bit CS technique were combined in~\cite{y1} and presented many advantages, e.g., the avoidance of uplink-bandwidth-resource occupation and the reduction of mutual interference, etc.
However, it is facing challenges in recovery accuracy and processing delay~\cite{a20}, etc.

By integrating the promising advantage of deep learning and inspired by the superimposed CSI feedback by using 1-bit CS in~\cite{y1}, we propose a DL-based 1-bit superimposed CSI feedback scheme in this paper. First, the downlink CSI is compressed by the 1-bit CS technique and then superimposed on the UL-US as feedback to the BS.
At the BS, to recover the bit information for both the UL-US and downlink CSI, a multi-task detection network with transmitted signal feature extraction is first constructed.
Then, with the recovered bits of the downlink CSI, a lightweight reconstruction network, which consists of an initial feature extraction of the downlink CSI with simplified traditional method and a single hidden layer network, is utilized to reconstruct the downlink CSI with a low processing delay.
Specifically, the advantages of superimposed CSI feedback by using 1-bit CS are inherited, i.e., without any occupation of uplink bandwidth for CSI feedback, and effective interference cancellation in~\cite{y1}, and the recovery accuracies for both the UL-US and downlink CSI are improved.

\subsection{Contributions}

In this paper, a DL-based 1-bit superimposed CSI feedback scheme is proposed to improve the superimposed CSI
feedback 1-bit CS approach in~\cite{y1}. To the best of our knowledge, there is a little literature focusing on the DL-based 1-bit superimposed CSI feedback method. And there is also no research on the introduction of deep learning into 1-bit superimposed feedback. The main contributions of this paper are as follows:
\begin{itemize}

\item We propose the DL-based scheme for 1-bit CS-based superimposed CSI feedback. By using the nonlinear mapping and feature extraction ability of the DL, we develop a detection network and a reconstruction network to further suppress nonlinear superimposition interference, and improve the detection and reconstruction performances. The proposed scheme retains the advantages of 1-bit CS-based superimposed CSI feedback~\cite{y1}, while obtains better recovery accuracy for both the UL-US and downlink CSI with much lower processing delay.

\item We construct a multi-task detection network to recover the bit information for both the UL-US and downlink CSI, based on the model-driven approach and assisted by the superimposition-interference cancellation technology. This detection network is jointly trained to detect the UL-US and downlink CSI simultaneously, capturing a globally optimized network parameter. We use the ability that DL solve nonlinear problems to solve the superimposition separation, which shortens processing delay while improving the detection performance without any second-order statistical information about channel and noise.
%

\item We develop a lightweight reconstruction network by using the linear approximation ability of the traditional superimposed coding aided binary iterative hard thresholding (SCA-BIHT) algorithm and the advantages of deep learning to deal with nonlinear problems. In this network, the initial feature of downlink CSI is extracted by SCA-BIHT algorithm with only a few iterations, and then a single hidden layer refinement network is constructed to refine the downlink CSI reconstruction. The reconstruction network not only greatly reduces the iterations of the traditional SCA-BIHT algorithm to raise efficiency, but also obtains a better reconstruction performance of the downlink CSI with a lower processing delay.

\end{itemize}

The remainder of this paper is structured as follows: In Section II, we introduce the system model of the 1-bit superimposed CSI feedback. The DL-based 1-bit superimposed CSI feedback method is presented in Section III and followed by numerical results in Section IV. Finally, Section V concludes our work.

Notations: Boldface upper case and lower case letters denote matrix and vector respectively. ${\left(\cdot \right)^T}$ and $\mathbf{(\cdot)^\dag}$ denote transpose and matrix pseudo-inverse respectively. $\mathbf{I}_P$ is the identity matrix of size $P \times P$. ${\mathrm {{\mathbb{BN}}}}\left( \cdot \right)$ denotes the operation of batch normalization. ${\left\|  \cdot  \right\|_2}$ is the Euclidean norm. ${\rm{sign}}(\cdot)$ denotes an operator of taking symbolic information, e.g., the sign function returns +1 for positive numbers and 0 otherwise. ${\rm{Re}}(\cdot)$ and ${\rm{Im}}(\cdot)$ represent real and imaginary part operations, respectively. $K(\mathbf{x})$ represents computing the best $k$-term approximation of $\mathbf{x}$ by thresholding. $\odot$ denotes the operation of Hadamard product for two vectors or matrices.

\section{System Model}

\begin{figure}[!h]
\includegraphics[width=3.5in]{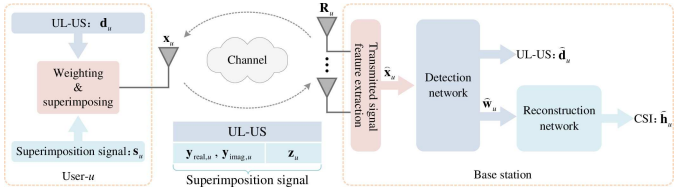}
\caption{\bf System model.\label{fig1}}
\end{figure}

The system model is shown in Fig~\ref{fig1}. Considering a massive MIMO system that consists of one BS with $N$ antennas and $U$ single-antenna users, after the processing of matched-filter, the received signal from user-$u$, $u = 1, 2, \ldots, U$, denoted as ${{\mathbf{R}}_{u}}$, is given as
\begin{eqnarray}
\label{EQ1}
{{\mathbf{R}}_u} = {{\mathbf{g}}_u}{{\mathbf{x}}_u} + {{\mathbf{N}}_u},
\end{eqnarray}
where ${{\mathbf{g}}_{u}}\in {{\mathbb{C}}^{N\times 1}}$ denotes the uplink channel vector from user-$u$ to the BS, ${{\mathbf{N}}_{u}}\in {{\mathbb{C}}^{N\times P}}$ is the circularly symmetric complex Gaussian noise (CSCG) of feedback link, $P$ is the length of the UL-US. To avoid occupying the limited and crowded uplink bandwidth resources~\cite{a33}\cite{a34}, $\mathbf{x}_u\in {{\mathbb{C}}^{1\times P}}$ adopts superimposition technology, and denotes the transmitted signal of user-$u$, which is given by~\cite{y1}
\begin{eqnarray}
\label{EQ2}
{{\mathbf{x}}_u} = \sqrt {{\rho {E_u}}} {{\mathbf{s}}_u} + \sqrt {(1 - \rho ){E_u}} {{\mathbf{d}}_u},
\end{eqnarray}
where ${{\rho }}\in \left[ 0,1 \right]$ is the power proportional coefficient of the downlink CSI, $E_u$ is the transmitted power of user-$u$, and ${{\mathbf{s}}_{u}}\in {{\mathbb{C}}^{1\times P}}$ and ${{\mathbf{d}}_{u}}\in {{\mathbb{C}}^{1\times P}}$ stand for the modulated superimposition signal and the UL-US, respectively.

In this paper, the downlink CSI, satisfying ${{\mathbf{h}}_u}\in {{\mathbb{C}}^{1\times N}}$, is a sparse vector with $K$-sparsity~\cite{y1}, i.e., only $K$ non-zero elements in ${{\mathbf{h}}_u}$. According to the 1-bit CS technique~\cite{a21}, ${{\mathbf{h}}_u}$ is compressed by
\begin{eqnarray}
\label{EQ3}
\left\{ {\begin{array}{*{20}{c}}
{{{\mathbf{y}}_{{\mathrm{real,}}u}} = \mathrm{sign}({\mathop{\rm Re}\nolimits} ({{\mathbf{h}}_u}){{\mathbf{\Phi }}_u})}\\
{{{\mathbf{y}}_{{\mathrm{imag,}}u}} = \mathrm{sign}({\mathop{\rm Im}\nolimits} ({{\mathbf{h}}_u}){{\mathbf{\Phi }}_u})}
\end{array}} \right.,
\end{eqnarray}
where ${{\mathbf{\Phi }}_u}\in {{\mathbb{R}}^{N\times M}}$ is the measurement matrix~\cite{y1}, and ${{\mathbf{y}}_{{\mathrm{real,}}u}}\in {{\mathbb{R}}^{1\times M}}$ and ${{\mathbf{y}}_{{\mathrm{imag,}}u}}\in {{\mathbb{R}}^{1\times M}}$ denote the real and imaginary parts of the compressed CSI, respectively.

For the convenience of digital modulation, the support-set of the downlink CSI ${{\mathbf{h}}_u}$, denoted as ${{\mathbf{z}}_u}\in{\{ 0,1\} ^{1 \times N}}$, is labelled by the bit-form~\cite{y1}, i.e.,
\begin{eqnarray}
\label{EQ201}
{z_{u,k}} = \left\{ \begin{array}{l}
\;1,\;\;{h_{u,k}} \ne 0\\
\,0,\;\;{h_{u,k}} = 0
\end{array}\right.\;,\;k = 1,2, \ldots N,
\end{eqnarray}
where ${z_{u,k}}$ and ${h_{u,k}}$ are the \textit{k}-th element in ${\mathbf{z}_u}$ and ${\mathbf{h}_u}$, respectively.
In order to reconstruct a more accurate downlink CSI at the BS, ${\bf{z}}_{u}$ needs to be fed back to the BS with ${{\mathbf{y}}_{{\mathrm{real,}}u}}$ and ${{\mathbf{y}}_{{\mathrm{imag,}}u}}$ by using the feedback vector $\mathbf{p}_u$. The feedback vector $\mathbf{p}_u$ is formed by merging ${{\mathbf{y}}_{{\mathrm{real,}}u}}$, ${{\mathbf{y}}_{{\mathrm{imag,}}u}}$, ${{\mathbf{z}}_u}$~\cite{y1}, i.e.,
\begin{eqnarray}
\label{EQ4}
{{\mathbf{p}}_u} = {[{{\mathbf{y}}_{{\mathrm{real,}}u}},{{\mathbf{y}}_{{\mathrm{imag,}}u}},{{\mathbf{z}}_u}]}.
\end{eqnarray}
It is worth noting that ${{\mathbf{p}}_u}$ can be viewed as a bit stream with the elements of ${{\mathbf{p}}_u}$ only being 0 or 1. With the digital modulation, we have
\begin{eqnarray}
\label{EQ5}
{{\mathbf{w}}_u} \buildrel \Delta \over = f_{\mathrm{modu}} ( { {\mathbf{p}}_u}),
\end{eqnarray}
where $f_{\mathrm{modu}}    \left(\cdot \right)$ denotes the mapping function of digital modulation, such as the quadrature phase shift keying (QPSK). In Eq~(\ref{EQ5}), ${{\mathbf{p}}_u}$ is mapped as \textit{modulated feedback vector} (MFV) ${{\mathbf{w}}_u}\in {{\mathbb{C}}^{1\times L}}$, where ${\textit{L}} = \lceil \left( {2M + N } \right)/2 \rceil$. Without loss of generality, the UL-US's length $P$ is larger than $L$ due to main task of user services~\cite{b1}\cite{b2}. Similar to~\cite{y1} and~\cite{b2}, to superimpose MVF with UL-US, a spread spectrum method is utilized, which could capture spread spectrum gain to suppress the interference caused by the superimposition processing. Thus, the superimposition signal ${{\mathbf{s}}_u}$, given in Eq~(\ref{EQ2}), is obtained by using a spreading matrix to spread the MFV ${{\mathbf{w}}_u}$, i.e.,
\begin{eqnarray}
\label{EQ6}
{{\mathbf{s}}_u} = \frac{1}{\sqrt L}{{{{\mathbf{w}}_u}{{\mathbf{Q}}_u}}},
\end{eqnarray}
where ${{\mathbf{Q}}_u} \in {{\mathbb{R}}^{L \times P}}$ is a spreading matrix, which satisfies ${{\mathbf{Q}}_u}{\mathbf{Q}}_u^T = P{{\mathbf{I}}_L}$, e.g., the Walsh matrix~\cite{a35}. By combining Eq~(\ref{EQ2}) and Eq~(\ref{EQ6}), the transmitted signal of user-$u$ ${{\mathbf{x}}_u}$ is rewritten as
\begin{eqnarray}
\label{EQ7}
{{\mathbf{x}}_u} = \sqrt {\frac{{\rho {E_u}}}{L}} {{\mathbf{w}}_u}{{\mathbf{Q}}_u} + \sqrt {(1 - \rho ){E_u}} {{\mathbf{d}}_u}.
\end{eqnarray}

At the user-$u$, the downlink CSI ${{\mathbf{h}}_u}$ is compressed by using 1-bit CS (given in Eq~(\ref{EQ3})), and thus the transmitted signal ${{\mathbf{x}}_u}$ is formed by weighting and superimposing the UL-US ${{\mathbf{d}}_u}$ and superimposition signal ${{\mathbf{s}}_u}$ according to Eq~(\ref{EQ2})--(\ref{EQ7}). With the received ${\mathbf{R}}_{u}$ at the BS, the \textit{detection network} and \textit{reconstruction network} are designed to detect the UL-US ${{\mathbf{d}}_u}$ and superimposition signal ${{\mathbf{s}}_u}$, and recover the downlink CSI ${{\mathbf{h}}_u}$, respectively. The detection and reconstruction networks will be deliberated in Section III.

\section{DL-based Superimposed CSI Feedback Using 1-Bit CS}
In this section, according to the superimposed CSI feedback scheme with the 1-bit CS\cite{y1}, the \textit{detection network} and \textit{reconstruction network} are developed to recover the UL-US and downlink CSI. A transmitted signal feature extraction is first employed to coarsely extract the feature after equalizing the uplink wireless channel. Then, with the extracted transmitted signal feature, we design the detection network and reconstruction network.

\subsection{Transmitted Signal Feature Extraction}

From Eq~(\ref{EQ2})--(\ref{EQ7}), the transmitted signal ${{\mathbf{x}}_u}$ is formed by superimposing the UL-US ${{\mathbf{d}}_u}$ and the modulated superimposition signal ${{\mathbf{s}}_u}$.
To recover ${{\mathbf{d}}_u}$ and ${{\mathbf{s}}_u}$, the transmitted signal ${{\mathbf{x}}_u}$ should be first extracted, and thus the uplink channel ${{\mathbf{g}}_u}$ in Eq~(\ref{EQ1}) needs to be removed by channel equalization.
From~\cite{y1} and~\cite{b1}, the transmitted signal feature extraction is employed in this paper. That is, the uplink wireless channel is equalized through zero forcing (ZF) equalization, so as to extract the transmission signal feature. The feature extraction is given as
\begin{eqnarray}\label{EQ8}
{{\mathbf{\mathord{\buildrel{\lower3pt\hbox{$\scriptscriptstyle\frown$}}
\over x} }}_u} = {\mathbf{g}}_u^\dag {{\mathbf{R}}_u} = {{\mathbf{x}}_u} + {\mathbf{g}}_u^\dag {{\mathbf{N}}_u},
\end{eqnarray}
where ${{\mathbf{\mathord{\buildrel{\lower3pt\hbox{$\scriptscriptstyle\frown$}}
\over x} }}_u}$ denotes the coarse extracted vector of transmitted signal ${{\mathbf{x}}_u}$.
It should be noted that, relative to the use of ZF equalization to extract the transmitted signal feature, the use of minimum mean square error (MMSE) channel equalization can obtain better feature extraction performance, while encounters higher computational complexity. Especially, the MMSE equalization requires second-order statistics of uplink channel $\mathbf{g}_u$ and noise $\mathbf{N}_u$~\cite{y1}\cite{a12}, which leads to application difficulties. Therefore, we use low-complexity ZF equalization to extract the transmitted signal feature, leaving the feature improvement to the subsequent detection network.

With the extracted transmitted signal feature ${{\mathbf{\mathord{\buildrel{\lower3pt\hbox{$\scriptscriptstyle\frown$}}
\over x} }}_u}$, we construct the detection network to detect UL-US ${{\mathbf{d}}_u}$ and superimposition signal ${{\mathbf{s}}_u}$. From Eq~(\ref{EQ6}), ${{\mathbf{s}}_u}$ is obtained by spreading the MFV ${{\mathbf{w}}_u}$. In addition, the compressed downlink CSI ${{\mathbf{y}}_{{\mathrm{real,}}u}}$ and ${{\mathbf{y}}_{{\mathrm{imag,}}u}}$ can be recovered from ${{\mathbf{w}}_u}$ (given in Eq~(\ref{EQ3})--(\ref{EQ5})).

\subsection{Detection Network}
In order to eliminate superimposed interference and obtain better downlink CSI and UL-US reconstruction accuracy, the detection network is designed by using unfolding method \cite{a36}. That is, the iteration steps in \cite{y1} are replaced by the groups of CSI-Net and Det-Net, including six subnets, i.e., CSI-Net1, Det-Net1, CSI-Net2, Det-Net2, CSI-Net3, and Det-Net3, in which the UL-US ${{\mathbf{d}}_u}$ and MFV ${{\mathbf{w}}_u}$ are detected by solving a multi-task problem.

\subsubsection{Architecture}

\begin{figure}[!h]
\centering
\includegraphics[width=3.5in]{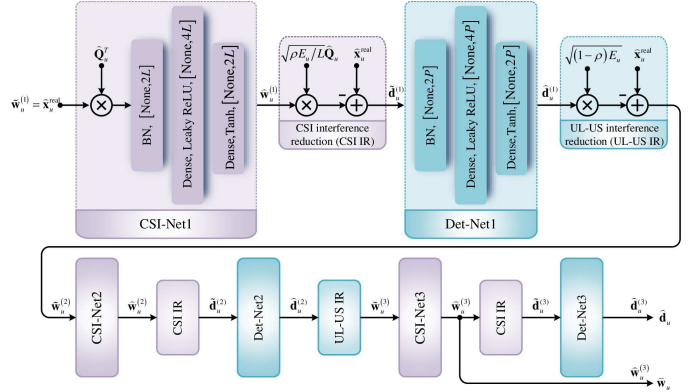}
\caption{\bf Detection network}
\label{fig2}
\end{figure}

The architecture of detection network is illustrated in Fig~\ref{fig2}. From the perspective of convenience and ease of implementation, we first use the easiest single hidden layer neural network architecture to design CSI-Net$i$ and Det-Net$i$ ($i=1,2,3$). After experimental verification, this architecture is not only easy but also improves performance. The architecture of detection network is described as follows:

\begin{itemize}
\item CSI-Net1, DET-Net1, CSI-Net2, DET-Net2, CSI-Net3, and DET-Net3 are successively cascaded to form the multi-task network. To reduce mutual interference, some expert knowledge is inserted between each cascaded subnets, i.e., the interference cancellation technology~\cite{a12}\cite{b1}. In more detail, the CSI interference reduction (CSI IR) is introduced between the CSI-Net$i$ and Det-Net$i$ ($i=1,2,3$), while the UL-US interference reduction (UL-US IR) is inserted between Det-Net$i$ and CSI-Net$(i+1)$ ($i=1,2$).

\item The same network structures are employed by the CSI-Net$i$ and Det-Net$i$ ($i=1,2,3$). Each subnet consists of an input layer, a hidden layer, and an output layer with a fully connected mode. For each CSI-Net$i$ (DET-Net$i$) ($i = 1,2,3$), the number of neurons in the input layer, hidden layer, and output layer are $2L$ ($2P$), $4L$ ($4P$), and $2L$ ($2P$), respectively.

\item For each subnet, a batch normalization (BN) is employed to normalize its input sets, converting the subnet input to zero mean and unit variance.

\item The activation functions of linear activation, leaky rectified linear unit (LReLU)~\cite{a23} and hyperbolic tangent (Tanh) are adopted by the input layer, hidden layer and output layer of each subnet, respectively.

\item The outputs of CSI-Net3 and DET-Net3 are the detected MFV ${\mathbf{\mathord{\buildrel{\lower3pt\hbox{$\scriptscriptstyle\frown$}}
\over w} }}_u$ (${\mathbf{\mathord{\buildrel{\lower3pt\hbox{$\scriptscriptstyle\frown$}}
\over w} }}_u = {\mathbf{\mathord{\buildrel{\lower3pt\hbox{$\scriptscriptstyle\frown$}}
\over w} }}_u^{\left( 3 \right)}$) and the detected UL-US ${\mathbf{\mathord{\buildrel{\lower3pt\hbox{$\scriptscriptstyle\frown$}}
\over d} }}_u$ (${\mathbf{\mathord{\buildrel{\lower3pt\hbox{$\scriptscriptstyle\frown$}}
\over d} }}_u = {\mathbf{\mathord{\buildrel{\lower3pt\hbox{$\scriptscriptstyle\frown$}}
\over d} }}_u^{\left( 3 \right)}$), respectively.

\end{itemize}

The network architecture is summarized in Table~\ref{table_I}.

\begin{table}[!ht]

\renewcommand{\arraystretch}{1.2}
\caption{\bf Architecture of detection network}
\label{table_I}
\centering

\begin{tabu}{|c|[1pt]c|c|c|c|c|c|}
\hline
Subnet     & \multicolumn{3}{c|}{CSI-Net$i$}& \multicolumn{3}{c|}{Det-Net$i$} \\ \tabucline[1pt]{-}
layer      & Input   & Hidden   & Output   & Input   & Hidden   & Output   \\ \hline
BN         & $\surd$ & $\times$ &$\times$  & $\surd$ & $\times$ & $\times$ \\ \hline
Neurons    & 2$L$      & 4$L$       & 2$L$       & 2$P$      & 4$P$       & 2$P$       \\ \hline
Activation & None    & LReLU    & Tanh     & None    & LReLU    & Tanh     \\ \hline
\end{tabu}
\end{table}

\subsubsection{Process of detection network}

\begin{itemize}
  \item \textit{Data preprocessing}
\end{itemize}

Due to the requirement of real-valued data sets in common DL-based framework, we transform the coarse extracted complex-valued vectors ${{\mathbf{\mathord{\buildrel{\lower3pt\hbox{$\scriptscriptstyle\frown$}}\over x} }}_u}$, UL-US ${{\mathbf{d}}_u}$ and MFV ${{\mathbf{w}}_u}$ to the real-valued vectors, i.e.,
\begin{eqnarray}\label{EQ9}
\left\{ {\begin{array}{*{20}{c}}
{{\mathbf{\mathord{\buildrel{\lower3pt\hbox{$\scriptscriptstyle\frown$}}
\over x} }}_u^{\mathrm{real}} = \left[ {{\rm{Re}}({{{\mathbf{\mathord{\buildrel{\lower3pt\hbox{$\scriptscriptstyle\frown$}}
\over x} }}}_u}),{\rm{Im}}({{{\mathbf{\mathord{\buildrel{\lower3pt\hbox{$\scriptscriptstyle\frown$}}
\over x} }}}_u})} \right]}\\
{{\mathbf{d}}_u^{\mathrm{real}} = \left[ {{\rm{Re}}({{\mathbf{d}}_u}),{\rm{Im}}({{\mathbf{d}}_u})} \right]}\;\;\;\\
{{\mathbf{w}}_u^{\mathrm{real}} = \left[ {{\rm{Re}}({{\mathbf{w}}_u}),{\rm{Im}}({{\mathbf{w}}_u})} \right]}\;\;\;
\end{array}} \right..
\end{eqnarray}

To match the real-valued operation, the spreading matrix ${{\mathbf{Q}}_u}$ is also transformed to real-valued matrix ${{\mathbf{\mathord{\buildrel{\lower3pt\hbox{$\scriptscriptstyle\frown$}}
\over Q} }}_u}$, which is obtained as
\begin{eqnarray}\label{EQ10}
{{\mathbf{\mathord{\buildrel{\lower3pt\hbox{$\scriptscriptstyle\frown$}}
\over Q} }}_u} = \left[ {\begin{array}{*{20}{c}}
{{{\mathbf{Q}}_u}}&\mathbf{0}\\
\mathbf{0}&{{{\mathbf{Q}}_u}}
\end{array}} \right].
\end{eqnarray}

Then, to train the detection network, ${\mathbf{\mathord{\buildrel{\lower3pt\hbox{$\scriptscriptstyle\frown$}}
\over x} }}_u^{\mathrm{real}}$ is employed as the network input, while ${\mathbf{d}}_u^{\mathrm{real}}$ and ${\mathbf{w}}_u^{\mathrm{real}}$ are used as training labels in the CSI-Net$i$ and Det-Net$i$ ($i=1,2,3$), respectively.
In addition, to facilitate the unified description of the sub-network input in the detection network, we use ${\mathbf{\widetilde w}}_u^{(1)}$ to represent the input ${\mathbf{\mathord{\buildrel{\lower3pt\hbox{$\scriptscriptstyle\frown$}}
\over x} }}_u^{\mathrm{real}}$ of the detection network, i.e., ${\mathbf{\widetilde w}}_u^{(1)}={\mathbf{\mathord{\buildrel{\lower3pt\hbox{$\scriptscriptstyle\frown$}}\over x} }}_u^{real}$.

\begin{itemize}
  \item \textit{Processing procedure}
\end{itemize}

The processing procedure of trained detection network is given in Table~\ref{table_II}, and some steps are explained as follows.

\begin{table}[!ht]
\centering
\renewcommand\arraystretch{1.2}
\caption{\bf Processing procedure}
\label{table_II}
\begin{tabular}{l}
\hline
\hline

\textbf{Input:} ${\mathbf{\widetilde w}}_u^{(1)}={\mathbf{\mathord{\buildrel{\lower3pt\hbox{$\scriptscriptstyle\frown$}}
\over x} }}_u^{real}$. \\
\hline

\kern 6pt (1-1): Use CSI-Net1 to detect MFV $\mathbf{w}_u$, then capture ${\mathbf{\mathord{\buildrel{\lower3pt\hbox{$\scriptscriptstyle\frown$}}
\over w} }}_u^{\left( 1 \right)}$.\\

\kern 6pt (1-2): Perform CSI IR by using expert knowledge and acquire ${\mathbf{\tilde d}}_u^{\left( 1 \right)}$.\\

\kern 6pt (1-3): Detect UL-US by using DET-Net1 to obtain ${\mathbf{\mathord{\buildrel{\lower3pt\hbox{$\scriptscriptstyle\frown$}}
\over d} }}_u^{\left( 1 \right)}$.\\

\kern 6pt (1-4): Use expert knowledge to carry out UL-US IR and get ${\mathbf{\tilde w}}_u^{\left( 1 \right)}$. \\

\kern 6pt (2-1): Employ CSI-Net2 to detect MFV and capture ${\mathbf{\mathord{\buildrel{\lower3pt\hbox{$\scriptscriptstyle\frown$}}
\over w} }}_u^{\left( 2 \right)}$.\\

\kern 6pt (2-2): Conduct CSI IR by using expert knowledge and acquire ${\mathbf{\tilde d}}_u^{\left( 2 \right)}$.\\

\kern 6pt (2-3): Detect UL-US by using DET-Net2 to obtain ${\mathbf{\mathord{\buildrel{\lower3pt\hbox{$\scriptscriptstyle\frown$}}
\over d} }}_u^{\left( 2 \right)}$.\\

\kern 6pt (2-4):  Perform UL-US IR by using expert knowledge and get ${\mathbf{\tilde w}}_u^{\left( 2 \right)}$. \\

\kern 6pt (3-1): Use CSI-Net3 to detect MFV, then capture ${\mathbf{\mathord{\buildrel{\lower3pt\hbox{$\scriptscriptstyle\frown$}}
\over w} }}_u^{\left( 3 \right)}$.\\

\kern 6pt (3-2): Use expert knowledge to conduct CSI IR and acquire ${\mathbf{\tilde d}}_u^{\left( 3 \right)}$. \\

\kern 6pt (3-3): Detect UL-US by using DET-Net3 to obtain ${\mathbf{\mathord{\buildrel{\lower3pt\hbox{$\scriptscriptstyle\frown$}}
\over d} }}_u^{\left( 3 \right)}$.\\

\hline

\textbf{Output:} ${\mathbf{\mathord{\buildrel{\lower3pt\hbox{$\scriptscriptstyle\frown$}}
\over w} }}_u = {\mathbf{\mathord{\buildrel{\lower3pt\hbox{$\scriptscriptstyle\frown$}}
\over w} }}_u^{\left( 3 \right)}$, and ${\mathbf{\mathord{\buildrel{\lower3pt\hbox{$\scriptscriptstyle\frown$}}
\over d} }}_u = {\mathbf{\mathord{\buildrel{\lower3pt\hbox{$\scriptscriptstyle\frown$}}
\over d} }}_u^{\left( 3 \right)}$.\\

 \hline
 \hline
\end{tabular}
\end{table}

\textbf{Process of CSI-Net$i$}: The CSI-Net$i$ ($i=1,2,3$), is used to detect the MFV, which is expressed as
\begin{small}
\begin{eqnarray}\label{EQ11}
{\mathbf{\mathord{\buildrel{\lower3pt\hbox{$\scriptscriptstyle\frown$}}
\over w} }}_u^{(i)} = \sigma_2\left( {{\mathbf{W}}_{12}^{(i)}\left( {\sigma_1\left( {{\mathbf{W}}_{11}^{(i)}\mathbb{BN}\left( {{\mathbf{\widetilde w}}_u^{(i)}{\mathbf{\mathord{\buildrel{\lower3pt\hbox{$\scriptscriptstyle\frown$}}
\over Q} }}_u^T} \right){\rm{ + }}{\mathbf{b}}_{11}^{(i)}} \right)} \right){\rm{ + }}{\mathbf{b}}_{12}^{(i)}} \right),
\end{eqnarray}
\end{small}
\noindent where $\sigma_1$ and $\sigma_2$ denote the activation functions of LReLU and Tanh, respectively. In Eq~(\ref{EQ11}), ${\mathbf{W}}_{11}^{(i)}$ (${\mathbf{b}}_{11}^{(i)}$) and ${\mathbf{W}}_{12}^{(i)}$ (${\mathbf{b}}_{12}^{(i)})$ are the weights (biases) of hidden layer and output layer, respectively. We use CSI-Net$i$ to detect MFV $\mathbf{w}_u$ and obtain the network output ${\mathbf{\mathord{\buildrel{\lower3pt\hbox{$\scriptscriptstyle\frown$}}
\over w} }}_u^{(i)}$, which is briefly described in steps (1-1), (2-1), and (3-1) in Table~\ref{table_II}.

\textbf{CSI IR}:
In steps (1-2), (2-2), and (3-2) in Table~\ref{table_II}, to reduce the interference from MFV, a spreading is employed by CSI IR, which is expressed as
\begin{eqnarray}\label{EQ12}
{\mathbf{\widetilde d}}_u^{(i)} = {\mathbf{\mathord{\buildrel{\lower3pt\hbox{$\scriptscriptstyle\frown$}}
\over x} }}_u^{\mathrm{real}} - \sqrt {\frac{{\rho {E_u}}}{L}}  {\mathbf{\mathord{\buildrel{\lower3pt\hbox{$\scriptscriptstyle\frown$}}
\over w} }}_u^{\left( i \right)} {{\mathbf{\mathord{\buildrel{\lower3pt\hbox{$\scriptscriptstyle\frown$}}
\over Q} }}_u},
\end{eqnarray}
where ${{\mathbf{\mathord{\buildrel{\lower3pt\hbox{$\scriptscriptstyle\frown$}}
\over Q} }}_u}$ is obtained according to Eq~(\ref{EQ10}). Then, ${\mathbf{\widetilde d}}_u^{(i)}$ is fed into Det-Net$i$ to detect the UL-US.

\textbf{Process of Det-Net$i$}: The Det-Net$i$ ($i=1,2,3$), is used to detect the UL-US, which is expressed as
\begin{eqnarray}\label{EQ13}
{\mathbf{\mathord{\buildrel{\lower3pt\hbox{$\scriptscriptstyle\frown$}}
\over d} }}_u^{(i)} = \sigma_2\left( {{\mathbf{W}}_{22}^{(i)}\left( {\sigma_1\left( {{\mathbf{W}}_{21}^{(i)}\mathbb{BN}\left( {{\mathbf{\widetilde d}}_u^{(i)}} \right){\rm{ + }}{\mathbf{b}}_{21}^{(i)}} \right)} \right){\rm{ + }}{\mathbf{b}}_{22}^{(i)}} \right),
\end{eqnarray}
where ${\mathbf{W}}_{21}^{(i)}$ (${\mathbf{b}}_{21}^{(i)}$) and ${\mathbf{W}}_{22}^{(i)}$ (${\mathbf{b}}_{22}^{(i)})$ denote the weights (biases) of hidden layer and output layer, respectively.

\textbf{UL-US IR}: In steps (1-4) and (2-4) in Table~\ref{table_II},  to reduce the interference from the UL-US, the outputs of Det-Net$i$ ($i=1,2,3$) are processed by expert knowledge, which is expressed as
\begin{eqnarray}\label{EQ14}
{\mathbf{\widetilde w}}_u^{(i)} = {\mathbf{\mathord{\buildrel{\lower3pt\hbox{$\scriptscriptstyle\frown$}}
\over x} }}_u^{\mathrm{real}} - \sqrt {(1 - \rho ){E_u}} {\mathbf{\mathord{\buildrel{\lower3pt\hbox{$\scriptscriptstyle\frown$}}
\over d} }}_u^{(i)}.
\end{eqnarray}

With the process given in Table~\ref{table_II}, the UL-US ${{\mathbf{d}}_u}$ and MFV ${{\mathbf{w}}_u}$ are detected, where the real-valued descriptions of the detected UL-US ${{\mathbf{d}}_u}$ and MFV ${{\mathbf{w}}_u}$ are denoted by ${\mathbf{\mathord{\buildrel{\lower3pt\hbox{$\scriptscriptstyle\frown$}}
\over d} }}_u$ and ${\mathbf{\mathord{\buildrel{\lower3pt\hbox{$\scriptscriptstyle\frown$}}
\over w} }}_u$, respectively. Then, with the detected MFV  ${\mathbf{\mathord{\buildrel{\lower3pt\hbox{$\scriptscriptstyle\frown$}}
\over w} }}_u$, we develop the reconstruction network to recover the downlink CSI ${{\mathbf{h}}_u}$.

\subsection{Reconstruction Network}

\begin{figure}[!h]
\centering
\includegraphics[width=3.5in]{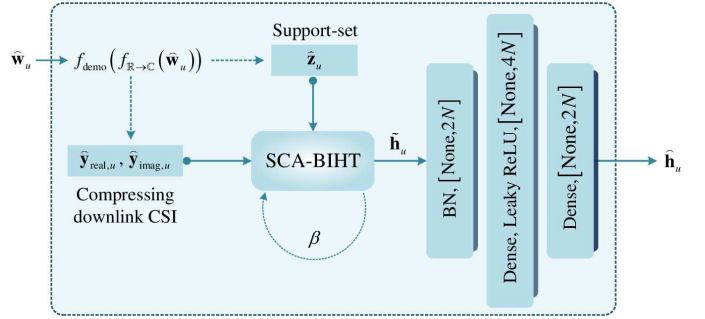}
\caption{\bf Reconstruction network}
\label{fig3}
\end{figure}

\begin{table}[!ht]
\centering
\renewcommand\arraystretch{1.2}
\caption{\bf Procedure of reconstruction network}
\label{table_III}
\begin{tabular}{l}
\hline
\hline

\textbf{Input:} ${\mathbf{\mathord{\buildrel{\lower3pt\hbox{$\scriptscriptstyle\frown$}}
\over w} }}_u$.\\
\hline

\kern 6pt 1):  Map the real-valued ${\mathbf{\mathord{\buildrel{\lower3pt\hbox{$\scriptscriptstyle\frown$}}
\over w} }}_u$ to $[{{\mathbf{\mathord{\buildrel{\lower3pt\hbox{$\scriptscriptstyle\frown$}}
\over y} }}_{{\mathrm{real,}}u}},{{\mathbf{\mathord{\buildrel{\lower3pt\hbox{$\scriptscriptstyle\frown$}}
\over y} }}_{{\mathrm{imag,}}u}},{{\mathbf{\mathord{\buildrel{\lower3pt\hbox{$\scriptscriptstyle\frown$}}
\over z} }}_u}]$\\

\kern 6pt 2): Use $[{{\mathbf{\mathord{\buildrel{\lower3pt\hbox{$\scriptscriptstyle\frown$}}
\over y} }}_{{\mathrm{real,}}u}},{{\mathbf{\mathord{\buildrel{\lower3pt\hbox{$\scriptscriptstyle\frown$}}
\over y} }}_{{\mathrm{imag,}}u}},{{\mathbf{\mathord{\buildrel{\lower3pt\hbox{$\scriptscriptstyle\frown$}}
\over z} }}_u}]$ to rough extract the feature of downlink \\ \kern 17pt CSI and
obtain ${{\mathbf{\tilde h}}_u}$ by the SCA-BIHT algorithm with $\beta$ times of iteration.\\

\kern 6pt 3): Use the refinement network to refine the reconstructed downlink CSI ${{\mathbf{\mathord{\buildrel{\lower3pt\hbox{$\scriptscriptstyle\frown$}}
\over h} }}_u}$. \\

\hline

\textbf{Output:} ${{\mathbf{\mathord{\buildrel{\lower3pt\hbox{$\scriptscriptstyle\frown$}}
\over h} }}_u}$. \\

 \hline
 \hline
\end{tabular}
\end{table}

A reconstruction network is designed to further improve the reconstruction accuracy of ${{\mathbf{h}}_u}$ on the basis of the reconstruction algorithm, and to reduce the processing delay caused by multiple iterations of the reconstruction algorithm.
The reconstruction network is given in Fig~\ref{fig3}, and the processing procedure is summarized in Table~\ref{table_III}. Generally, the corresponding de-mapping is first employed to restore the compressed downlink CSI. Then, the reconstruction algorithm given in~\cite{y1} with reduced complexity is utilized to perform an initial feature extraction of the downlink CSI. According to the initial feature extraction, two dense layers are used to refine the reconstruction of the downlink CSI. These details will be presented as follows.

\subsubsection{Inverse mapping operation}

From Eq~(\ref{EQ5}) and (\ref{EQ9}), the real-valued ${\mathbf{w}}_u^{\mathrm{real}}$ is formed by digital modulation and the mapping from complex-valued to real-valued form. Correspondingly, we adopt inverse mapping to recover the complex-valued and unmodulated forms. An inverse mapping, denoted by $f_{\mathbb{R} \rightarrow \mathbb{C}}(\cdot)$, is first employed to map the real-valued ${\mathbf{\mathord{\buildrel{\lower3pt\hbox{$\scriptscriptstyle\frown$}}
\over w} }}_u$ back to its complex-valued form. Then, the digital demodulation mapping, denoted as $f_{{\mathrm{demo}}}(\cdot)$, is used to demodulate this complex-valued vector. The whole inverse mapping process is expressed as
\begin{eqnarray}\label{EQ15}
[{{\mathbf{\mathord{\buildrel{\lower3pt\hbox{$\scriptscriptstyle\frown$}}
\over y} }}_{{\mathrm{real,}}u}},{{\mathbf{\mathord{\buildrel{\lower3pt\hbox{$\scriptscriptstyle\frown$}}
\over y} }}_{{\mathrm{imag,}}u}},{{\mathbf{\mathord{\buildrel{\lower3pt\hbox{$\scriptscriptstyle\frown$}}
\over z} }}_u}] \buildrel \Delta \over = {f_{{\mathrm{demo}}}}  \left ( f_{\mathbb{R} \rightarrow \mathbb{C}} \left ({\mathbf{\mathord{\buildrel{\lower3pt\hbox{$\scriptscriptstyle\frown$}}
\over w} }}_u \right) \right).
\end{eqnarray}
Then, the estimation of sparsity $K$ of the downlink CSI, denoted as ${{{\mathord{\buildrel{\lower3pt\hbox{$\scriptscriptstyle\frown$}}
\over K} }}}$, is obtained by calculating the number of non-zero entries in ${{\mathbf{\mathord{\buildrel{\lower3pt\hbox{$\scriptscriptstyle\frown$}}
\over z} }}_u}$.

\subsubsection{Initial feature extraction}

\begin{table}[!ht]
\centering
\renewcommand\arraystretch{1.2}
\caption{\bf Initial feature extraction procedure}
\label{table_IIII}
\begin{tabular}{l}
\hline
\hline

\textbf{Input:} Compressed downlink CSI $[{{\mathbf{\mathord{\buildrel{\lower3pt\hbox{$\scriptscriptstyle\frown$}}
\over y} }}_{{\mathrm{real,}}u}},{{\mathbf{\mathord{\buildrel{\lower3pt\hbox{$\scriptscriptstyle\frown$}}
\over y} }}_{{\mathrm{imag,}}u}}]$, Support-set ${{\mathbf{\mathord{\buildrel{\lower3pt\hbox{$\scriptscriptstyle\frown$}}
\over z} }}_u}$. \\
\hline

\kern 6pt 1): \textbf{Initialization:} the real part and imaginary part of reconstructed \\
\kern 16pt data are set to zero (i.e., ${\mathbf{h}}_{\mathrm{real}}^0 = {\mathbf{h}}_{\mathrm{imag}}^0 = {\mathbf{0}}$), $t=0$, and \\
\kern 16pt maximum number of iterations $\beta$.\\

\kern 6pt 2): $t = t+1$. \\

\kern 6pt 3): ${\mathbf{h}}_{\mathrm{real}}^t = K({\mathbf{h}}_{\mathrm{real}}^{t - 1} + ({{\mathbf{\mathord{\buildrel{\lower3pt\hbox{$\scriptscriptstyle\frown$}}
\over y} }}_{{\mathrm{real,}}u}} - \mathrm{sign}({\mathbf{h}}_{\mathrm{real}}^{t - 1}{{\mathbf{\Phi }}_u})){\mathbf{\Phi }}_u^T)$,\\
\kern 17pt  ${\mathbf{h}}_{\mathrm{imag}}^t = K({\mathbf{h}}_{\mathrm{imag}}^{t - 1} + ({{\mathbf{\mathord{\buildrel{\lower3pt\hbox{$\scriptscriptstyle\frown$}}
\over y} }}_{{\mathrm{imag,}}u}} - \mathrm{sign}({\mathbf{h}}_{\mathrm{imag}}^{t - 1}{{\mathbf{\Phi }}_u})){\mathbf{\Phi }}_u^T)$.\\

\kern 6pt 4): ${\mathbf{h}}_{\mathrm{real}}^t = {\mathbf{h}}_{\mathrm{real}}^t \odot {{\mathbf{\mathord{\buildrel{\lower3pt\hbox{$\scriptscriptstyle\frown$}}
\over z} }}_u},{\mathbf{h}}_{\mathrm{imag}}^t = {\mathbf{h}}_{\mathrm{imag}}^t \odot {{\mathbf{\mathord{\buildrel{\lower3pt\hbox{$\scriptscriptstyle\frown$}}
\over z} }}_u}$.\\

\kern 6pt 5): Go to step 2) if $t < \beta$, else go to next step. \\

\kern 6pt 6):  ${\mathbf{h}} = {\mathbf{h}}_{\mathrm{real}}^t + j{\mathbf{h}}_{\mathrm{imag}}^t$.\\

\kern 6pt 7): Normalization: ${{\mathbf{\tilde h}}_u} = {\mathbf{h}}/{\left\| {\mathbf{h}} \right\|_2}$.\\

\hline

\textbf{Output:} ${{\mathbf{\tilde h}}_u}$. \\

 \hline
 \hline
\end{tabular}
\end{table}

With ${{\mathbf{\mathord{\buildrel{\lower3pt\hbox{$\scriptscriptstyle\frown$}}
\over y} }}_{{\mathrm{real,}}u}}$, ${{\mathbf{\mathord{\buildrel{\lower3pt\hbox{$\scriptscriptstyle\frown$}}
\over y} }}_{{\mathrm{imag,}}u}}$, ${{\mathbf{\mathord{\buildrel{\lower3pt\hbox{$\scriptscriptstyle\frown$}}
\over z} }}_u}$ and ${{{\mathord{\buildrel{\lower3pt\hbox{$\scriptscriptstyle\frown$}}
\over K} }}}$, we employ the reconstruction algorithm, named SCA-BIHT in~\cite{b1}, to conduct an initial feature extraction of the downlink CSI, while leaving the refinement reconstruction to the subsequent refinement network. In particular, this initial feature extraction is executed by SCA-BIHT with only a few iterations instead of dozens or hundreds of iterations in~\cite{y1}. Here, $\beta$ times of iteration are adopted in this paper. The initial feature extraction procedure is presented in Table~\ref{table_IIII}.

Based on the initial feature extraction, we then input ${{\mathbf{\tilde h}}_u}$ to a single hidden layer network to refine the reconstruction accuracy of the downlink CSI $\mathbf{h}_u$.

\subsubsection{Refinement network}

\begin{table}[]

\renewcommand{\arraystretch}{1.2}
\caption{\bf Architecture of refinement network}
\label{table_V}
\centering

\begin{tabu}{|c|[1pt]c|c|c|}
\hline
Layer               & Input & Hidden & Output \\ \tabucline[1pt]{-}
Batch normalization &$\surd$            & $\times$     &$\times$      \\ \hline
Neuron number       & 2$N$    & 4$N$     & 2$N$     \\ \hline
Activation function & None  & LReLU  & Linear   \\ \hline
\end{tabu}
\end{table}

According to the initial feature extraction, a single hidden layer network is employed to refine the reconstruction of the downlink CSI, and its network architecture is summarized in Table~\ref{table_V}. Similar to CSI-Net$i$ and Det-Net$i$ ($i=1,2,3$) of the detection network, the refinement network is also designed as the easiest single hidden layer neural network architecture.

The the initial feature of downlink CSI ${{\mathbf{\tilde h}}_u}$ and the label ${{\mathbf{h}}_{u}}$ are complex-valued, and thus need to be mapped to real-valued form, i.e.,
\begin{eqnarray}\label{EQ16}
\left\{ \begin{gathered}
  {\mathbf{\tilde h}}_u^{{\text{real}}} = \left[ {\operatorname{Re} \left( {{{{\mathbf{\tilde h}}}_u}} \right),\operatorname{Im} \left( {{{{\mathbf{\tilde h}}}_u}} \right)} \right] \hfill \\
  {\mathbf{h}}_u^{{\text{real}}} = \left[ {\operatorname{Re} \left( {{{\mathbf{h}}_u}} \right),\operatorname{Im} \left( {{{\mathbf{h}}_u}} \right)} \right] \hfill \\
\end{gathered}  \right..
\end{eqnarray}
Then, using the refinement network, the refined reconstruction of the downlink CSI is obtained from the expression
\begin{eqnarray}\label{EQ17}
{{\mathbf{\mathord{\buildrel{\lower3pt\hbox{$\scriptscriptstyle\frown$}}
\over h} }}_u} = {{\mathbf{W}}_{32}}\left( {\sigma_1\left( {{{\mathbf{W}}_{31}}\mathbb{BN}\left( {\mathbf{\tilde h}}_u^{{\text{real}}} \right){\rm{ + }}{{\mathbf{b}}_{31}}} \right)} \right){\rm{ + }}{{\mathbf{b}}_{32}},
\end{eqnarray}
where ${{\mathbf{W}}_{31}}$ (${{\mathbf{b}}_{31}}$) and ${{\mathbf{W}}_{32}}$ (${{\mathbf{b}}_{32}}$) denote weights (biases) of the hidden layer and output layer of the refinement network, respectively.

\subsection{Model Training Specification}

Since model training is significant for network performance, we give the training details in this subsection. In the following, we discuss the training method, data preparation, and loss function, respectively.

\subsubsection*{Training method}

In this paper, the detection network and reconstruction network are separately trained to reduce the complexity of parameter tuning. For detection network, there are six subnetworks needed to be trained, including the training parameters ${\mathbf{W}}_{11}^{(i)}$, ${\mathbf{W}}_{12}^{(i)}$, ${\mathbf{W}}_{21}^{(i)}$, ${\mathbf{W}}_{22}^{(i)}$, ${\mathbf{b}}_{11}^{(i)}$, ${\mathbf{b}}_{12}^{(i)}$, ${\mathbf{b}}_{21}^{(i)}$, and ${\mathbf{b}}_{22}^{(i)}$ ($i = 1,2,3$). From Fig~\ref{fig2}, the detection network is a multi-task network in reality, which generates the estimated MFV ${\mathbf{\mathord{\buildrel{\lower3pt\hbox{$\scriptscriptstyle\frown$}}
\over w} }}_u$ and UL-US ${\mathbf{\mathord{\buildrel{\lower3pt\hbox{$\scriptscriptstyle\frown$}}
\over d} }}_u$, respectively. Thus, we jointly train the six subnets of detection network to resolve this multi-task issue. In the reconstruction network, only the refinement network needs to be trained to optimize its network parameters ${{\mathbf{W}}_{31}}$, ${{\mathbf{W}}_{32}}$, ${{\mathbf{b}}_{31}}$, and ${{\mathbf{b}}_{32}}$. With the trained detection network and the corresponding initial feature extraction of reconstruction network, we then train the refinement network solely.

\subsubsection{Data preparation for training}
The training set is acquired by a simulation approach, in which a significant amount of data samples are generated to train two networks, i.e., the detection network and the refinement network. Specially, these data samples are generated as follows.

${{\mathbf{h}}_u}$ and ${{\mathbf{g}}_u}$ are randomly generated on the basis of the distribution $\mathcal{CN}\left( {0,\left( {{1 \mathord{\left/{\vphantom {1 N}} \right. \kern-\nulldelimiterspace} N}} \right)} \right)$. To train the detection network, we first collect the ${{\mathbf{\mathord{\buildrel{\lower3pt\hbox{$\scriptscriptstyle\frown$}}\over x} }}_u}$ according to Eq~(\ref{EQ8}) to form input sets. Then we save the corresponding ${{\mathbf{d}}_{u}}$ and ${{\mathbf{w}}_u}$ as target sets, where ${{\mathbf{d}}_{u}}$ is formed by QPSK modulation with randomly generated Bernoulli binary sequences. All the complex-valued data sets are converted to real-valued form.
For example, the input and label of the detection network are set as $\left\{ {\left( {{\mathbf{\mathord{\buildrel{\lower3pt\hbox{$\scriptscriptstyle\frown$}}
\over x} }}_u^{\mathrm{real}}} \right),\left( {{\mathbf{d}}_u^{\mathrm{real}},{\mathbf{w}}_u^{\mathrm{real}}} \right)} \right\}$ according to Eq~(\ref{EQ9}).
Similarly, the input and label of the refinement network are set as $\left\{ {\left( {\mathbf{\tilde h}}_u^{{\mathrm{real}}} \right),\left( {{\mathbf{h}}_u^{\mathrm{real}}} \right)} \right\}$ according to Eq~(\ref{EQ16}).
In addition, to validate the trained network parameters during the training phase, a validation set is generated by following the same generation method of training set, and thus we could capture a set of optimized network parameters.

\subsubsection{Loss functions}

\begin{figure}[!b]
\centering
\includegraphics[width=3.5in]{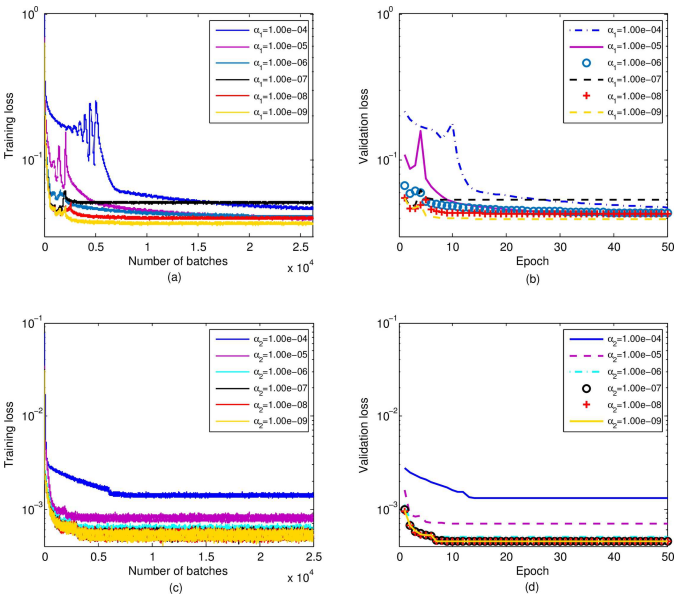}
\caption{(a) Training loss of the detection network. (b) Validation loss of the detection network. (c) Training loss of the reconstruction network. (d) Validation loss of the reconstruction network.}\label{fig4}
\end{figure}

The detection network is trained by optimizing weights and biases of each subnet, i.e., CSI-Net$i$ and Det-Net$i$, to minimize the loss function~\cite{a24}\cite{a25}.
In addition, the $l_2$ regularization is employed in the detection network to avoid gradient explosions~\cite{a26}.
Thus, the loss function for training the detection network is expressed as
\begin{eqnarray}\label{EQ18}
{\rm{Los}}{{\rm{s}}_{{\rm{\texttt{Det}}}}}{\rm{ = los}}{{\rm{s}}_{\rm{1}}} + \alpha_1 {\left\| {\mathbf{\Theta}_1}  \right\|_2^2},
\end{eqnarray}
where $\alpha_1$ is the regularization coefficient and ${\mathbf{\Theta}_1}$ denotes the training parameters, i.e., weights and biases of the detection network.
In Eq~(\ref{EQ18}), $\mathrm{loss_1}$ represents the weighted sum of the losses of six subnets, which is given as
\begin{eqnarray}\label{EQ19}
{\rm{loss_1 = }}\frac{1}{6}\left (\sum\limits_{i{\rm{ = }}1}^3 {\left\| {{\mathbf{\mathord{\buildrel{\lower3pt\hbox{$\scriptscriptstyle\frown$}}
\over d} }}_u^{(i)} - {\mathbf{d}}_u^{real}} \right\|_2^2}  + \sum\limits_{i{\rm{ = }}1}^3 {\left\| {{\mathbf{\mathord{\buildrel{\lower3pt\hbox{$\scriptscriptstyle\frown$}}
\over w} }}_u^{(i)} - {\mathbf{w}}_u^{\mathrm{real}}} \right\|_2^2}\right ),
\end{eqnarray}
where ${\mathbf{\mathord{\buildrel{\lower3pt\hbox{$\scriptscriptstyle\frown$}}
\over d} }}_u^{(i)}$ and ${\mathbf{\mathord{\buildrel{\lower3pt\hbox{$\scriptscriptstyle\frown$}}
\over w} }}_u^{(i)}$ are the output of the CSI-Net$i$ and Det-Net$i$, respectively. With this detection network, we obtain the MFV ${\mathbf{\mathord{\buildrel{\lower3pt\hbox{$\scriptscriptstyle\frown$}}
\over w} }}_u^{\left( 3 \right)}$ and UL-US ${\mathbf{\mathord{\buildrel{\lower3pt\hbox{$\scriptscriptstyle\frown$}}
\over d} }}_u^{\left( 3 \right)}$, i.e., ${\mathbf{\mathord{\buildrel{\lower3pt\hbox{$\scriptscriptstyle\frown$}}
\over w} }}_u$ and ${\mathbf{\mathord{\buildrel{\lower3pt\hbox{$\scriptscriptstyle\frown$}}
\over d} }}_u$.

With the trained detection network, the reconstruction network is trained according to ${{\mathbf{\mathord{\buildrel{\lower3pt\hbox{$\scriptscriptstyle\frown$}}
\over y} }}_{{\mathrm{real,}}u}}$, ${{\mathbf{\mathord{\buildrel{\lower3pt\hbox{$\scriptscriptstyle\frown$}}
\over y} }}_{{\mathrm{imag,}}u}}$, and ${{\mathbf{\mathord{\buildrel{\lower3pt\hbox{$\scriptscriptstyle\frown$}}
\over z} }}_u}$, which are detected by the detection network and expressed in Eq~(\ref{EQ15}). In reconstruction network, only the refinement network with single hidden layer needs to be optimized, and thus the loss function is given by
\begin{eqnarray}\label{EQ20}
{\rm{Los}}{{\rm{s}}_{\texttt{Rec}}} = \left\| {{{{\mathbf{\mathord{\buildrel{\lower3pt\hbox{$\scriptscriptstyle\frown$}}
\over h} }}}_u} - {\mathbf{h}}_u^{\mathrm{real}}} \right\|_2^2 + {\alpha _2}{\left\| {{{\mathbf{\Theta }}_2}} \right\|_2^2},
\end{eqnarray}
where ${{{\mathbf{\mathord{\buildrel{\lower3pt\hbox{$\scriptscriptstyle\frown$}}
\over h} }}}_u}$ is the estimated downlink CSI, $\alpha _2$ is the regularization coefficient and ${\mathbf{\Theta }}_2$ denotes all training parameters of refinement network.

To reap an effective and feasible regularization coefficient and verify the generalization performance of detection network and reconstruction network, Fig~\ref{fig4} compares the convergence behaviors of ${\rm{Los}}{{\rm{s}}_{\texttt{Det}}}$ and ${\rm{Los}}{{\rm{s}}_{\texttt{Rec}}}$ under different regularization coefficients (i.e., $\alpha _{1/2} = 10^{-9},10^{-8},..., 10^{-4}$). From Fig~\ref{fig4}, we can observe the convergence values of training loss and validation loss are almost the same, which indicates the excellent generalization performance of detection and reconstruction network. In addition, a smaller value of $\alpha _1$ (or $\alpha _2$) leads to a smaller convergence value of training loss or validation loss. Yet according to Eq~(\ref{EQ20}), the value of ${\rm{Los}}{{\rm{s}}_{\texttt{Rec}}}$ is related to  $\alpha _2$, the $\alpha _2$ that minimizes the  ${\rm{Los}}{{\rm{s}}_{\texttt{Rec}}}$ may not achieve the best reconstruction performance. The optimized $\alpha _2$ is determined by the reconstruction performance of the downlink CSI, which will be given in the experimental analysis.

By using the trained detection network and reconstruction network, the UL-US ${{{\mathbf{\mathord{\buildrel{\lower3pt\hbox{$\scriptscriptstyle\frown$}}
\over d} }}}_u} $ and downlink CSI ${{{\mathbf{\mathord{\buildrel{\lower3pt\hbox{$\scriptscriptstyle\frown$}}
\over h} }}}_u}$ can be recovered from the proposed scheme. Compared with the 1-bit CS-based superimposed CSI feedback scheme in~\cite{y1}, both the recoveries of the UL-US and downlink CSI are improved by the proposed scheme, while the requirements of second-order statistics of noise are avoided. Besides, these improvements are robust against parameter variations, which will be presented in the experimental analysis.

\section{Experiment Results}

In this section, we give numerical results of the proposed scheme. Definitions and basic parameters involved in simulations are first given. Then, to verify the effectiveness of the proposed scheme, we show the bit error rate (BER) of UL-US and MFV, and the normalized mean squared error (NMSE) of reconstructed downlink CSI is presented. Finally, we compare the online running time between the proposed scheme and conventional scheme. The source code is available at https://github.com/qingchj851/DL-1BitCS-SC-CSI-Feedback2.

\subsection{Parameter Setting}

Definitions involved in simulations are given as follows. The signal-to-noise ratio (SNR) in decibel (dB) of the signal received at BS from user-$u$ is defined as \cite{b1}
\begin{equation}\label{EQ21}
\mathrm{SNR} = 10{\log _{10}}\left( {\frac{{{E_u}}}{{\sigma _u^2}}} \right).
\end{equation}
The NMSE is utilized to evaluate the recovery performance of downlink CSI, and defined as \cite{b1}
\begin{equation}\label{EQ22}
\mathrm{NMSE} = {\frac{{\left\| {{{{\mathbf{\mathord{\buildrel{\lower3pt\hbox{$\scriptscriptstyle\frown$}}
\over h} }}}_u} - {{\mathbf{h}}_u^{\textrm{real}}}} \right\|_2^2}}{{\left\| {{\mathbf{h}}_u^{\textrm{real}}} \right\|_2^2}}} .
\end{equation}
In the experiment phase, $P=512$, $N=64$, and the sampling rate $c$ is defined as $c=M/N$.
The measurement matrix is randomly generated and obeys the Gaussian distribution~\cite{a32}, and it is guaranteed that its row vector and the column vector of the compressed signal cannot be sparsely represented by each other.
The Walsh matrix generated by the Walsh sequence is employed as the spreading matrix ${{\mathbf{Q}}_u}$ \cite{a35}.
The UL-US ${\mathbf{d}}_u$ is formed by applying QPSK modulation upon randomly generated Bernoulli binary sequences.
The training input data-sets are generated according to Eq~(\ref{EQ1})--(\ref{EQ8}). Trainings of detection network and reconstruction network are carried out under the noise-free setting, and this is different from the training of the DL-based network in \cite{b1}, where the training SNR is set as 5dB. Testing data-sets are generated by using the same method as the training data-sets.
The sizes of training set, validation set and testing set of detection network are 60,000, 20,000, and 20,000, respectively. For the reconstruction network, 45,000, 15,000, and 15,000 samples are respectively employed for the training, validation, and testing. Both in detection network and reconstruction network, we use Adam optimizer as the training optimization algorithm, and the values of epoch and learning rate are set to 50 and 0.001, respectively.
In the simulations, we stop the testing for BER performance when at least 1000-bit errors are observed \cite{b1}\cite{b2}.

\begin{figure}[h]
\centering
\includegraphics[width=3in]{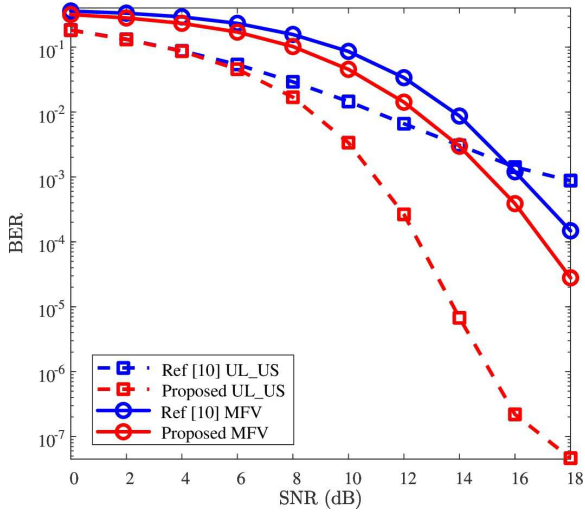}
\caption{BER versus SNR, where $P{\rm{ = 512}}$, $c = 2.0$, and $\rho = 0.10$ are considered.}\label{fig5}
\end{figure}

For the convenience of expression, we utilize ``Proposed'' and ``Ref\cite{y1}'' to denote the proposed DL-based 1-bit superimposed CSI feedback and the traditional 1-bit superimposed feedback (mentioned in \cite{y1}), respectively.

\subsection{BER Performance}

\begin{figure}[h]
\centering
\includegraphics[width=3in]{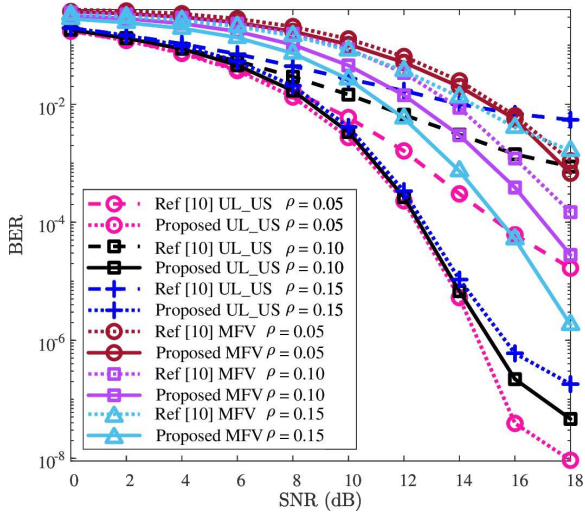}
\caption{BER versus SNR, where $P{\rm{ = 512}}$ and $c = 2.0$ are considered.}\label{fig6}
\end{figure}

\begin{figure}[h]
\centering
\includegraphics[width=3in]{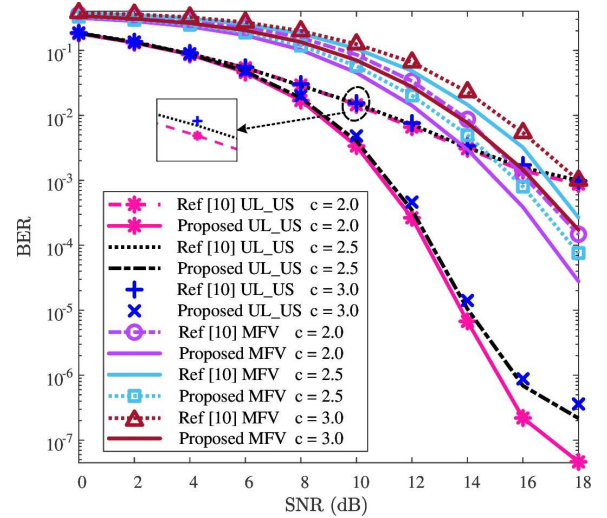}
\caption{BER versus SNR, where $P{\rm{ = 512}}$ and $\rho = 0.10$ are considered.}\label{fig7}
\end{figure}

In this subsection, the effectiveness and robustness of the detection network will be verified. To clarify the effectiveness, the comparison of BER's performance between ``Proposed'' and ``Ref\cite{y1}'' is first presented in Fig~\ref{fig5}. Next, to verify the robustness of the detection network, the impacts against the parameters of $\rho$ and $c$ are given in Fig~\ref{fig6} and Fig~\ref{fig7}, respectively.

To verify the effectiveness of the detection network, both the UL-US and MFV's BER performances are illustrated due to the UL-US being superimposed with MFV.
Fig~\ref{fig5} depicts the BER curves of the UL-US and MFV in terms of SNR, where $c=2.0$ and $\rho=0.10$ are considered.
From Fig~\ref{fig5}, the BERs of UL-US and MFV obtained by ``Proposed'' are respectively smaller than those of ``Ref\cite{y1}'' in the whole given SNR regions.
For example, when $\mathrm{SNR}=10$dB, the BER of UL-US (or MFV) by ``Proposed'' is around $3.4\times 10^{-3}$ (or $4.5 \times 10^{-2}$), while the BER of UL-US (or MFV) of ``Ref\cite{y1}'' is nearly $1.4\times 10^{-2}$ (or $8.5\times 10^{-2}$).
That is, compared with ``Ref\cite{y1}'', both the UL-US and MFV's BERs are improved by the proposed detection network.
Especially, these improvements are significant to be observed in the relatively higher SNR. The possible reason is that the detection network is trained under noise-free setting.

To verify the robustness of BER performance's improvement against the impact of $\rho$, the BER curves with different values of $\rho$, i.e., $\rho=0.05$, $\rho=0.10$, and $\rho=0.15$, are plotted in Fig~\ref{fig6}, where $c=2.0$ is considered. From Fig~\ref{fig6}, for each given $\rho$, the UL-US and MFV's BERs of the ``Proposed'' are respectively smaller than those of the ``Ref\cite{y1}''. This reflects that the proposed detection network could improve the BER performance under different $\rho$ for both UL-US and MFV. As $\rho$ increases from 0.05 to 0.15 for ``Proposed'', the BER of UL-US increases while the BER of MFV decreases, and vice versa. The reason is that the increased (or decreased) $\rho$ aggregates (or alleviates) the interference of MFV to UL-US, while alleviates (or aggregates) the interference of UL-US to MFV. On the whole, with the impacts of different $\rho$, the improvements of UL-US and MFV's BER performances are evidently observed. Thus, the proposed detection network guarantees the improvement of BER performance against the impact of $\rho$.

The UL-US and MFV's BER curves with different values of compression rate $c$ (i.e., $c=2.0$, $c=2.5$, and $c=3.0$) are depicted in Fig~\ref{fig7}, and this validates the improvement of BER performance is robust against the impact of $c$, where $\rho=0.10$. In Fig~\ref{fig7}, for each given $c$, the UL-US and MFV's BER performances of the ``Proposed'' are smaller than those of the ``Ref\cite{y1}''. This implies that the proposed detection network could improve UL-US and MFV's BER performance of ``Ref\cite{y1}'' for different values of $c$. With the increase of $c$, for both ``Proposed'' and ``Ref\cite{y1}'', the BERs of both UL-US and MFV increase, and vice versa. The reason is that the spreading gain (i.e., $P/M$) decreases with the increase of $c$, and thus affects the detection performances (similar results can be found in \cite{b1} and \cite{b2}). As a whole, compared with ``Ref\cite{y1}'', the BER improvements of UL-US and MFV are evidently observed for each given $c$. Thus, the proposed detection network shows its robustness of improving UL-US and MFV's BER performances against the impact of $c$.

\begin{figure}[h]
\centering
\includegraphics[width=3in]{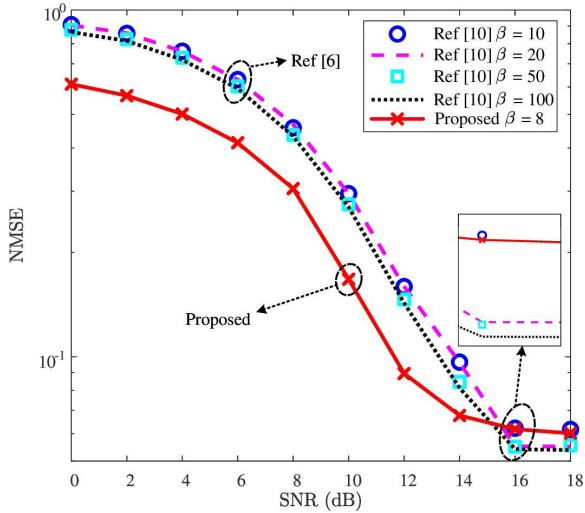}
\caption{NMSE versus SNR, where $P{\rm{ = 512}}$, $c = 2.0$, and $\rho = 0.10$ are considered.}\label{fig8}
\end{figure}

To sum up, according to Figs~\ref{fig5} - \ref{fig7}, the UL-US and MFV's BER performances of ``Ref\cite{y1}'' are effectively improved by the proposed detection network, and these improvements are robust against the impacts of $\rho$ and $c$.

\subsection{NMSE Performance}

\begin{figure}[h]
\centering
\includegraphics[width=3in]{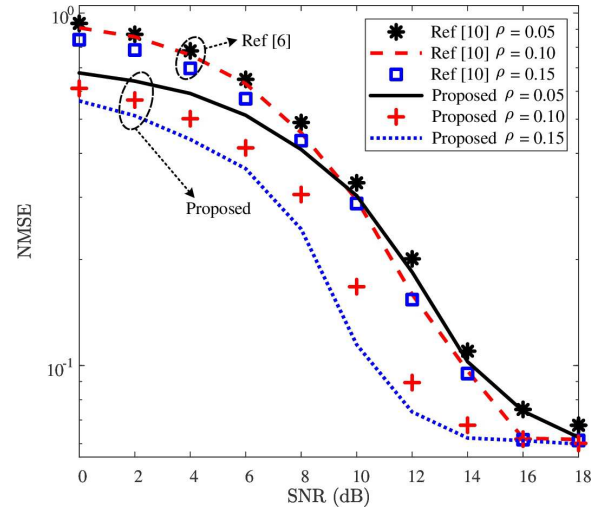}
\caption{NMSE versus SNR, where $P{\rm{ = 512}}$ and $c = 2.0$ are considered, and the $\beta$ of Ref\cite{y1} is 10.}\label{fig9}
\end{figure}

\begin{figure}[h]
\centering
\includegraphics[width=3in]{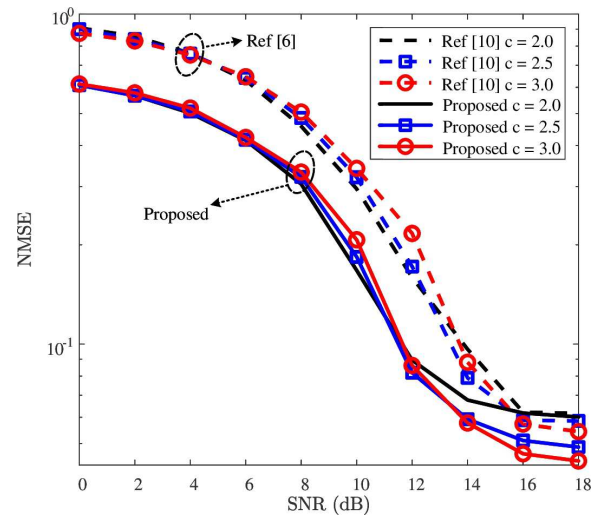}
\caption{NMSE versus SNR, where $P{\rm{ = 512}}$ and $\rho = 0.10$ are considered, and the $\beta$ of Ref\cite{y1} is 10.}\label{fig10}
\end{figure}

With the detected MFV, the downlink CSI can be reconstructed by using the proposed reconstruction network. To validate the effectiveness of the proposed reconstruction network, NMSE curves of the downlink CSI recovered from the proposed reconstruction network and SCA-BIHT \cite{y1} are first given in Fig~\ref{fig8}. Then, to demonstrate the robustness of the reconstruction network, the NMSE performance against the impacts of $\rho$ and $c$ are shown in Figs~\ref{fig9} and \ref{fig10}, respectively. In addition, we present the influence of regularization coefficient $\alpha_2$ on the NMSE performance in Table~\ref{table_VI}.

In Fig~\ref{fig8}, the NMSE curves of downlink CSI's recovery are depicted, where $c=2.0$ and $\rho=0.10$. The ``Proposed'' employs 8 times of iteration for initial feature extraction, i.e., $\beta = 8$, followed by two dense layers. In contrast, different iteration values (i.e., $\beta = 10$, $\beta = 20$, $\beta = 50$, and $\beta = 100$) are given for the SCA-BIHT algorithm of ``Ref\cite{y1}''. From Fig~\ref{fig8}, when $SNR\leq14$dB, the ``Proposed'' achieves the minimum NMSE, even lower than that of ``Ref\cite{y1}'' with $\beta = 100$. For example, when $\mathrm{SNR}=12$dB, the NMSE of ``Proposed'' is about $8.94\times 10^{-2}$, while that of ``Ref\cite{y1}'' with $\beta = 100$ is about $1.43\times 10^{-1}$. That is, with a smaller NMSE, the two dense layers in the reconstruction network can replace 95 iterations of SCA-BIHT algorithm in the relatively low SNR region (e.g., $\mathrm{SNR} \leq $14dB), leading to a lower processing delay. For the case where $\mathrm{SNR} \geq $16dB, the NMSE of ``Proposed'' outperforms that of ``Ref\cite{y1}'' with $\beta = 10$. Although it shows a slightly higher NMSE of ``Proposed'' than ``Ref\cite{y1}'' with $\beta = 50$ and $100$, it compensates the high processing delay of ``Ref\cite{y1}''. On the whole, the proposed reconstruction network has a lower processing delay than ``Ref\cite{y1}'' and shows a better NMSE performance in the relatively low SNR region. Therefore, the proposed reconstruction network is effective to improve the NMSE performance of ``Ref\cite{y1}''.

To verify the robust improvement of NMSE performance against the impact of $\rho$, the NMSE curves with variant $\rho$ (i.e., $\rho=0.05$, $\rho=0.10$, and $\rho=0.15$) are plotted in Fig~\ref{fig9}. From Fig~\ref{fig9}, for each given $\rho$, the downlink CSI's NMSE of the ``Proposed'' is smaller than that of the ``Ref\cite{y1}''. With the increase of $\rho$ (increases from 0.05 to 0.15), the NMSE decreases for both ``Ref\cite{y1}'' and ``Proposed'', and vice versa. The reason is that the downlink CSI can obtain more transmission power with a larger value of $\rho$. In addition, with the increase of SNR, the curves gradually converge for the reason that the main influence of NMSE comes from the superimposed interference in a relatively high SNR region. On the whole, for each given value of $\rho$ in Fig~\ref{fig9}, the NMSE of ``Ref\cite{y1}'' is reduced by the ``Proposed'', especially in the relatively low SNR region (e.g., $\mathrm{SNR} \leq $14dB). Thus, the proposed reconstruction network possesses its robustness for improving the NMSE performance against the impact of $\rho$.

Fig~\ref{fig10} plots the NMSE curves of downlink CSI with different values of compression rate $c$ (i.e., $c=2.0$, $c=2.5$, and $c=3.0$) to validate the robustness of NMSE performance's improvement against the impact of $c$. In Fig~\ref{fig10}, for each given $c$, the downlink CSI's NMSE performance of the ``Proposed'' is smaller than that of the ``Ref\cite{y1}''. In addition, for $\mathrm{SNR}\leq10$dB, the NMSEs of ``Proposed'' increase as the increase of $c$. The possible reason is that the higher compression rate results in lower spreading gain (i.e., $P/M$). In the low SNR region, the main impact of NMSE performance comes from the noise interference and is limited by the low spread spectrum gain. Yet, the NMSE's convergence value of high compression rate is smaller than that of low compression rate. For example, for the cases where $c=2.0$, $c=2.5$, and $c=3.0$, the convergence values of ``Proposed'' NMSE are about $6.0 \times 10^{-2}$, $4.9 \times 10^{-2}$, and $4.4 \times 10^{-2}$, respectively. The possible reason is that the higher compression rate brings more reconstruction information in the high SNR region, where the noise interference almost disappeared. On the whole, for each given value of $c$ in Fig~\ref{fig10}, the NMSE of ``Ref\cite{y1}'' is reduced by the ``Proposed''. Thus, the proposed reconstruction network possesses its robustness for improving the NMSE performance against the impact of $c$.

\begin{table*}[]

\renewcommand{\arraystretch}{1.2}
\caption{\bf The effect of regularization coefficient $\alpha_2$ on NMSE performance}
\label{table_VI}
\centering
\scalebox{0.88}{
\begin{tabu}{|c|[1pt]p{1.58cm}<{\centering}|p{1.58cm}<{\centering}|p{1.58cm}<{\centering}|p{1.58cm}<{\centering}|p{1.58cm}<{\centering}|p{1.58cm}<{\centering}|p{1.58cm}<{\centering}}
\hline
\diagbox[width=11.5em,dir=SE]{SNR (dB)}{Regularization\\coefficient} &$\alpha_2=10^{-4}$& $\mathbf{\alpha_2=10^{-5}}$& $\alpha_2=10^{-6}$ &$\alpha_2=10^{-7}$& $\alpha_2=10^{-8}$& $\alpha_2=10^{-9}$ \\
\Xhline{1pt}
 0 &$0.6121$ & \textbf{0.6114} & 0.6115 & 0.6116 & 0.6117 & 0.6118    \\ \hline

 2 &$0.5675$ & \textbf{0.5666} & 0.5668 & 0.5670 & 0.5671 & 0.5671    \\ \hline

 4 &$0.5024$ & \textbf{0.5015} & 0.5018 &0.5021 & 0.5022 & 0.5422  \\ \hline

 6 &$0.4151$ & \textbf{0.4145} & 0.4146 &0.4148 & 0.4149 & 0.4149     \\ \hline

 8 &$0.3063$ & \textbf{0.3056} & 0.3058 & 0.3060 & 0.3060 & 0.3061   \\ \hline

 10 &$0.1682$ & \textbf{0.1673} & 0.1675 & 0.1678 & 0.1678 & 0.1679   \\ \hline

 12 &$0.0903$ & \textbf{0.0894} & 0.0897 &0.0900 & 0.0900 & 0.0901   \\ \hline

 14 &$0.0683$ & \textbf{0.0677} & 0.0678 &0.0680 & 0.0681 & 0.0681    \\ \hline

 16 &$0.0626$ & \textbf{0.0618} & 0.0619 &0.0623 & 0.0623 & 0.0624   \\ \hline

 18 &$0.0608$ & \textbf{0.0601} & 0.0602 &0.0605 & 0.0606  & 0.0606    \\ \hline

\end{tabu}}
\end{table*}
 In addition, the influence of regularization coefficient $\alpha_2$ on NMSE performance is given in Table~\ref{table_VI}, where $c=2.0$, $\rho=0.10$, and different values of $\alpha_2$ (i.e., $\alpha_2=10^{-4}$, $\alpha_2=10^{-5}$, $\alpha_2=10^{-6}$, $\alpha_2=10^{-7}$, $\alpha_2=10^{-8}$, and $\alpha_2=10^{-9}$) are considered. From Table~\ref{table_VI}, the influence of different values of the regularization coefficient on the NMSE is not very obvious. Despite all this, among the given values of $\alpha_2$, in all SNR regions, the minimum of NMSE is still observed as $\alpha_2=10^{-5}$. Thus, the NMSE performance in Table~\ref{table_VI} indicates $\alpha_2=10^{-5}$ is a preferable regularization coefficient.

To sum up, according to Figs~\ref{fig8} - \ref{fig10}, the downlink CSI's NMSE performance of ``Ref\cite{y1}'' is effectively improved by the proposed reconstruction network, and these improvements are robust against the impacts of $\rho$ and $c$.

\begin{figure}[h]
\centering
\includegraphics[width=3in]{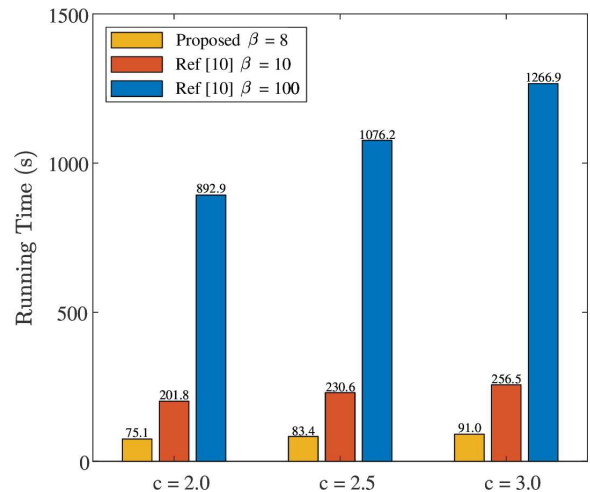}
\caption{Comparison of ``Proposed'' and ``Ref\cite{y1}'' about online running time by conducting $10^5$ times of experiments, where $P{\rm{ = 512}}$ and $\rho = 0.10$ are considered.}\label{fig11}
\end{figure}

\subsection{Online Running Time}
To illustrate the low processing delay of ``Proposed'', i.e., detection network and reconstruction network, the online running time between ``Proposed'' and ``Ref\cite{y1}'' is compared in Fig~\ref{fig11}, where $P{\rm{ = 512}}$, $\rho = 0.10$, and different values of $c$ (i.e., $c=2.0$, $c=2.5$, and $c=3.0$) are considered. Especially, ``Ref \cite{y1}'' adopts $\beta = 10$ and $100$ in the reconstruction algorithm (i.e., SCA-BIHT algorithm). Here, $\beta = 10$ in ``Ref\cite{y1}'' is used to guarantee the NMSE of the ``Proposed'' is smaller than that of ``Ref \cite{y1}'', and $\beta = 100$ in ``Ref\cite{y1}'' is used to present the ``Proposed'' has a similar NMSE (in a relatively high SNR region) while significantly lower processing delay as that of ``Ref\cite{y1}''. For a fair comparison, $10^5$ online-running experiments are conducted for ``Proposed'' and ``Ref\cite{y1}'' on the same PC (with CPU i5-8250U) by using MATLAB software. For each given $c$ in Fig~\ref{fig11}, the online running time of ``Proposed'' is shorter than that of ``Ref\cite{y1}'', e.g., when $c=2.0$, the online running time of ``Proposed'' and $\beta = 10$ ($\beta = 100$) in ``Ref\cite{y1}'' are 75.1s and 201.8s (1266.9s), respectively. This reflects that the proposed 1-bit CS-based superimposed CSI feedback can reduce the processing delay. It is also noticed that, as $c$ rises from 2.0 to 3.0, the online running time of both ``Proposed'' and ``Ref\cite{y1}'' go up. However, the total increased running time of the ``Proposed'' is 15.9s, which is far less than that of ``Ref\cite{y1}'' (e.g. 54.7s for $\beta = 10$ and 374.0s for $\beta = 100$). In addition, Fig~\ref{fig11} shows that the online running time of ``Ref\cite{y1}'' is proportional to the number of iteration. Thus, the NMSE performance might not be applicable for ``Ref\cite{y1}'' with large iteration number, while the ``Proposed'' can avoid this annoyance.

As a whole, compared with ``Ref\cite{y1}'', the proposed DL-based 1-bit superimposed CSI feedback significantly reduces the online running time.

\section{Conclusion}

The 1-bit CS-based superimposed CSI feedback is still facing many challenges, such as low recovery accuracy of the UL-US and downlink CSI, and long processing delay, etc. To remedy these defects, the DL-based 1-bit superimposed CSI feedback has been investigated in this paper. The constructed detection network captures optimized network parameters by using joint training, and thus improves the BER performance of the UL-US. Moreover, the detection network is also helpful for reconstructing the downlink CSI. With the detected downlink CSI's bits from the detection network, the proposed reconstruction network utilizes the simplified version of SCA-BIHT with a single hidden layer network, and achieves a significant improvement on NMSE performance of the downlink CSI recovery. In particular, compared with the conventional 1-bit CS-based superimposed CSI feedback, the proposed CSI feedback scheme presents its robustness against parameter variations and possesses significantly low processing delay.


\appendices

\ifCLASSOPTIONcaptionsoff
  \newpage
\fi


\begin{thebibliography}{00}
%
\bibitem{a1}
T.~Wu, X.~Yin, L.~Zhang, and J.~Ning, ``Measurement-based channel
  characterization for 5$\textrm{G}$ downlink based on passive sounding in
  $\textrm{S}$ub-6 $\textrm{G}$hz 5$\textrm{G}$ commercial network,''
  \emph{IEEE Trans. Wireless Commun.}, pp. 1--1, Jan.2021.

\bibitem{w1}
C.~Qing, W.~Yu, B.~Cai, J.~Wang, and C.~Huang, ``Elm-based frame
  synchronization in burst-mode communication systems with nonlinear
  distortion,'' \emph{IEEE Wireless Commun. Lett.}, vol.~9, no.~6, pp.
  915--919, June 2020.

\bibitem{z1}
Z. Wei, H. Li, H. Liu, B. Li, and C. Zhao, ``Randomized Low-Rank Approximation Based Massive MIMO CSI Compression,'' in \emph{IEEE Commun. Lett.}, vol. 25, no. 6, pp. 2004-2008, June 2021, doi: 10.1109/LCOMM.2021.3065751.

\bibitem{z2}
B. Lin, F. Gao, S. Zhang, T. Zhou, and A. Alkhateeb, ``Deep Learning-Based Antenna Selection and CSI Extrapolation in Massive MIMO Systems,'' in \emph{IEEE Trans. Wireless Commun.}, vol. 20, no. 11, pp. 7669-7681, Nov. 2021, doi: 10.1109/TWC.2021.3087318.

\bibitem{z3}
L. You, J. Wang, W. Wang, and X. Gao, ``Secure multicast transmission for massive MIMO with statistical channel state information,'' \emph{IEEE Signal Process. Lett.}, vol. 26, no. 6, pp. 803-807, June 2019.

\bibitem{z4}
X. Xia, K. Xu, S. Zhao, and Y. Wang, ``Learning the time-varying massive MIMO channels: Robust estimation and data-aided prediction,'' \emph{IEEE Trans. Veh. Technol.}, vol. 69, no. 8, pp. 8080-8096, Aug. 2020.

\bibitem{a2}
M.~{Sim}, J.~{Park}, C.~{Chae}, and R.~{Heath}, ``Compressed channel feedback
  for correlated massive $\textrm{MIMO}$ systems,'' \emph{J. Commun. Netw.},
  vol.~18, no.~1, pp. 95--104, Feb. 2016.

\bibitem{a3}
S.~Kim, J.~Choi, and B.~Shim, ``Feedback reduction for beyond 5$\textrm{G}$
  cellular systems,'' in \emph{Proc. IEEE Int. Conf. Commun. (ICC)}, May 2019,
  pp. 1--6.

\bibitem{a4}
W.~Shen, L.~Dai, Y.~Shi, X.~Zhu, and Z.~Wang, ``Compressive sensing based
  differential channel feedback for massive $\textrm{MIMO}$,'' \emph{Electron.
  Lett.}, vol.~51, no.~22, pp. 1824--1826, Oct. 2015.

\bibitem{y1}
C.~{Qing}, Q.~{Yang}, B.~{Cai}, B.~{Pan}, and J.~{Wang}, ``Superimposed
  coding-based $\textrm{CSI}$ feedback using 1-bit compressed sensing,''
  \emph{IEEE Commun. Lett.}, vol.~24, no.~1, pp. 193--197, Jan. 2020.

\bibitem{a5}
H.~{Son} and Y.~{Cho}, ``Analysis of compressed $\textrm{CSI}$ feedback in
  $\textrm{MISO}$ systems,'' \emph{IEEE Wireless Commun. Lett.}, vol.~8, no.~6,
  pp. 1671--1674, Aug. 2019.

\bibitem{a6}
P.~Wu, Z.~Liu, and J.~Cheng, ``Compressed $\textrm{CSI}$ feedback with learned
  measurement matrix for mm$\textrm{W}$ave massive $\textrm{MIMO}$,'' Jul 2020,
  $\textit{arXiv:1903.02127}$, [Online]. Available:
  https://arxiv.org/abs/1903.02127.

\bibitem{a7}
X.~{Rao} and V.~{Lau}, ``Distributed compressive $\textrm{CSI}$ estimation and
  feedback for $\textrm{FDD}$ multi-user massive $\textrm{MIMO}$ systems,''
  \emph{IEEE Trans. Signal Process.}, vol.~62, no.~12, pp. 3261--3271, June
  2014.

\bibitem{a8}
D.~{Jiang}, W.~{Wang}, L.~{Shi}, and H.~{Song}, ``A compressive sensing-based
  approach to end-to-end network traffic reconstruction,'' \emph{IEEE Trans.
  Netw. Sci. Eng.}, vol.~7, no.~1, pp. 507--519, Oct. 2020.

\bibitem{a9}
Z.~{Lu}, J.~{Wang}, and J.~{Song}, ``Multi-resolution $\textrm{CSI}$ feedback
  with deep learning in massive $\textrm{MIMO}$ system,'' in \emph{Proc. IEEE
  Int. Conf. Commun. (ICC)}, June 2020, pp. 1--6.

\bibitem{a10}
X.~{Li} and H.~{Wu}, ``Spatio-temporal representation with deep neural
  recurrent network in $\textrm{MIMO}$ $\textrm{CSI}$ feedback,'' \emph{IEEE
  Wireless Commun. Lett.}, vol.~9, no.~5, pp. 653--657, May 2020.

\bibitem{a11}
Q.~Sun, Y.~Wu, J.~Wang, C.~Xu, and K.~Wong, ``$\textrm{CNN}$ based
  $\textrm{CSI}$ acquisition for $\textrm{FDD}$ massive $\textrm{MIMO}$ with
  noisy feedback,'' \emph{Electron. Lett.}, vol.~55, no.~17, pp. 963--965, Jul.
  2019.

\bibitem{a12}
D.~{Xu}, Y.~{Huang}, and L.~{Yang}, ``Feedback of downlink channel state
  information based on superimposed coding,'' \emph{IEEE Commun. Lett.},
  vol.~11, no.~3, pp. 240--242, Mar. 2007.

\bibitem{b1}
C.~{Qing}, B.~{Cai}, Q.~{Yang}, J.~{Wang}, and C.~{Huang}, ``Deep learning for
  $\textrm{CSI}$ feedback based on superimposed coding,'' \emph{IEEE Access},
  vol.~7, pp. \textrm{93}\textrm{723}--\textrm{93}\textrm{733}, Jul. 2019.

\bibitem{b2}
C.~{Qing}, B.~{Cai}, Q.~{Yang}, J.~{Wang}, and C.~{Huang}, ``Elm-based
  superimposed $\textrm{CSI}$ feedback for $\textrm{FDD}$ massive
  $\textrm{MIMO}$ system,'' \emph{IEEE Access}, vol.~8, pp.
  \textrm{53}\textrm{408}--\textrm{53}\textrm{418}, Mar. 2020.

\bibitem{a13}
C.~Wen, W.~Shih, and S.~Jin, ``Deep learning for massive $\textrm{MIMO}$
  $\textrm{CSI}$ feedback,'' \emph{IEEE Wireless Commun. Lett.}, vol.~7, no.~5,
  pp. 748--751, Oct. 2018.

\bibitem{a14}
T.~{Wang}, C.~{Wen}, S.~{Jin}, and G.~{Li}, ``Deep learning-based
  $\textrm{CSI}$ feedback approach for time-varying massive $\textrm{MIMO}$
  channels,'' \emph{IEEE Wireless Commun. Lett.}, vol.~8, no.~2, pp. 416--419,
  Apr. 2019.

\bibitem{a15}
C.~{Lu}, W.~{Xu}, H.~{Shen}, J.~{Zhu}, and K.~{Wang}, ``$\textrm{MIMO}$ channel
  information feedback using deep recurrent network,'' \emph{IEEE Commun.
  Lett.}, vol.~23, no.~1, pp. 188--191, Jan. 2019.

\bibitem{a16}
Y.~Liao, H.~Yao, Y.~Hua, and C.~Li, ``$\textrm{CSI}$ feedback based on deep
  learning for massive $\textrm{MIMO}$ systems,'' \emph{IEEE Access}, vol.~7,
  pp. \textrm{86}\textrm{810}--\textrm{86}\textrm{820}, Jul. 2019.

\bibitem{a17}
T.~{Chen}, J.~{Guo}, S.~{Jin}, C.~{Wen}, and G.~{Li}, ``A novel quantization
  method for deep learning-based massive $\textrm{MIMO}$ $\textrm{CSI}$
  feedback,'' in \emph{Proc. IEEE Glob. Conf. Signal Inf. Process.}, Nov. 2019,
  pp. 1--5.

\bibitem{a18}
M.~{Mashhadi}, Q.~{Yang}, and
  D.~{G$\ddot{\mathrm{\textrm{u}}}$nd$\ddot{\mathrm{\textrm{u}}}$z},
  ``Distributed deep convolutional compression for massive $\textrm{MIMO}$
  $\textrm{CSI}$ feedback,'' \emph{IEEE Trans. Wireless Commun.}, vol.~20,
  no.~4, pp. 2621--2633, Apr. 2021.

\bibitem{a19}
W.~{Tang}, W.~{Xu}, X.~{Zhang}, and J.~{Lin}, ``A low-cost channel feedback
  scheme in mm$\textrm{W}$ave massive $\textrm{MIMO}$ system,'' in \emph{Proc.
  IEEE Int. Conf. Comput. Commun.}, Jul. 2017, pp. 89--93.

\bibitem{a20}
J.~{Wang}, Y.~{Ding}, S.~{Bian}, Y.~{Peng}, M.~{Liu}, and G.~{Gui},
  ``$\textrm{UL-CSI}$ data driven deep learning for predicting
  $\textrm{DL-CSI}$ in cellular $\textrm{FDD}$ systems,'' \emph{IEEE Access},
  vol.~7, pp. \textrm{96}\textrm{105}--\textrm{96}\textrm{112}, Jul. 2019.

\bibitem{a33}
Y Wu, L Qian, H Mao, W Lu, H Zhou, C Yu, ``Joint Channel Bandwidth and Power Allocations for Downlink
  Non-Orthogonal Multiple Access Systems,'' in \emph{Proc. IEEE Veh. Technol. Conf.}, pp. 1--5, Sep. 2019.


\bibitem{a34}
X Leturc, CJ Le~Martret, P Ciblat, ``Multiuser power and bandwidth allocation in ad hoc networks with
  Type-I HARQ under Rician channel with statistical CSI,'' in \emph{Proc. Int. Conf. Mil. Commun. Inf. Syst.}, pp. 1--7, May 2017.


\bibitem{a21}
P.~{Xiao}, B.~{Liao}, and J.~{Li}, ``One-bit compressive sensing via
  schur-concave function minimization,'' \emph{IEEE Trans. Signal Process.},
  vol.~67, no.~16, pp. 4139--4151, Aug. 2019.

\bibitem{a35}
A. Akansu and R. Poluri, ``Walsh-Like nonlinear phase orthogonal codes for direct sequence CDMA communications,'' \emph{IEEE Trans. Signal Process.}, vol.~55, no.~7, pp. 3800--3806, Jul. 2007, doi: 10.1109/TSP.2007.894229.

\bibitem{a36}
PK. Chundi, X. Wang, M. Seok, ``Channel Estimation Using Deep Learning on an
FPGA for 5G Millimeter-Wave Communication Systems,'' \emph{IEEE Trans. Circuits Syst. I-Regul. Pap.}, pp. 1--11, Oct. 2021.
doi:10.1109/TCSI.2021.3117886.

\bibitem{a23}
J.~{Guo}, C.~{Wen}, S.~{Jin}, and G.~{Li}, ``Convolutional neural network-based
  multiple-rate compressive sensing for massive $\textrm{MIMO}$ $\textrm{CSI}$
  feedback: design, simulation, and analysis,'' \emph{IEEE Trans. Wireless
  Commun.}, vol.~19, no.~4, pp. 2827--2840, Apr. 2020.

\bibitem{a24}
Y.~Zhang, X.~Wang, and H.~Tang, ``An improved elman neural network with
  piecewise weighted gradient for time series prediction,''
  \emph{Neurocomputing}, vol. 359, no.~24, pp. 199--208, Sep. 2019.

\bibitem{a25}
L.~Luo, Y.~Xiong, Y.~Liu, and X.~Sun, ``Adaptive gradient methods with dynamic
  bound of learning rate,'' Feb. 2019, $\textit{arXiv:1902.09843}$, [Online].
  Available: https://arxiv.org/abs/1902.09843.

\bibitem{a26}
M. Zeng, Y. Cai, X. Liu, Z. Cai, X. Li, ``Spectral-Spatial Clustering of Hyperspectral Image Based on Laplacian
  Regularized Deep Subspace Clustering,'' in
  \emph{Proc. IEEE Int. Geosci. Remote Sens. Symp.}, pp. 2694--2697, Jul. 2019.

\bibitem{a32}
F. Tong, L. Li, H. Peng, Y. Yang. ``Flexible construction of compressed sensing matrices with low storage
  space and low coherence,'' \emph{Signal Process.}, vol. 182, May 2021.


\end{thebibliography}
\end{document}